\documentclass[numberedappendix,iop]{emulateapj}

\usepackage{graphicx}
\usepackage{natbib}
\usepackage{color}
\usepackage[cmex10]{amsmath}

\shorttitle{Simulations of strong lensing}
\slugcomment{Preparing to submit to to ApJ: draft date \today}
\shortauthors{Nan Li et al.}

\begin{document}

\title{PICS: Simulations of Strong Gravitational Lensing in Galaxy Clusters}

\author{Nan Li\altaffilmark{1,2,3}; Michael D. Gladders\altaffilmark{1,3};
Esteban M. Rangel\altaffilmark{2,5}; Michael K. Florian\altaffilmark{1,3}; \\Lindsey E.
Bleem\altaffilmark{2,3}; Katrin Heitmann\altaffilmark{2,3}; Salman
Habib\altaffilmark{2,3}, and Patricia Fasel\altaffilmark{4}}

\altaffiltext{1}{Department of Astronomy \& Astrophysics, The University of
Chicago, 5640 South Ellis Avenue, Chicago, IL 60637, USA; {\em \email{linan7788626@oddjob.uchicago.edu}}}
\altaffiltext{2}{High Energy Physics Division, Argonne National Laboratory, Lemont, IL 60439, USA}
\altaffiltext{3}{Kavli Institute for Cosmological Physics at the University of
Chicago, 5640 South Ellis Avenue, Chicago, IL 60637, USA}
\altaffiltext{4}{Computer, Computational, and Statistical Sciences Division, Los
Alamos National Laboratory, Los Alamos, NM 87545, USA}
\altaffiltext{5}{Electrical Engineering and Computer Science, Northwestern
University, 2145 Sheridan Road, Evanston, IL 60208, USA}

\begin{abstract}

Gravitational lensing has become one of the most powerful tools
available for investigating the `dark side' of the
universe. Cosmological strong gravitational lensing, in particular,
probes the properties of the dense cores of dark matter halos over
decades in mass and offers the opportunity to study the distant universe
at flux levels and spatial resolutions otherwise unavailable. Studies of 
strongly-lensed variable sources offer yet further scientific
opportunities. One of the challenges in realizing the potential of
strong lensing is to understand the statistical context of both the individual systems 
that receive extensive follow-up study, as well as that of the
larger samples of strong lenses that are now emerging from survey
efforts. Motivated by these challenges, we have developed an image-simulation
pipeline, PICS (Pipeline for Images of Cosmological Strong lensing) to generate realistic strong gravitational lensing signals
from group- and cluster-scale lenses. PICS uses a low-noise and
unbiased density estimator based on (resampled) Delaunay Tessellations to
calculate the density field; lensed images are produced by
ray-tracing images of actual galaxies from deep Hubble Space Telescope
observations. Other galaxies, similarly sampled, are added to fill in the light cone. The pipeline further adds cluster-member galaxies and
foreground stars into the lensed images. The entire image ensemble is
then observed using a realistic point spread function which includes 
appropriate detector artifacts for bright stars. Noise is further added,
including such non-Gaussian elements as noise window-paning from
mosaiced observations, residual bad pixels, and cosmic rays. The aim is to produced simulated images that appear identical---to the
eye (expert or otherwise)---to real observations in various imaging
surveys. 

\end{abstract}


\section{Introduction}
\label{sec:introduction}

Gravitational lensing is, put simply, the deflection of photon paths
when they pass through a gravitational potential. In recent years,
gravitational lensing has come to the fore as a powerful tool to
investigate the ``dark side'' of the Universe \cite[for reviews, see
  e.g.,][and references therein]{Massey2010, Kneib2011, Treu2013, Meneghetti2013}.  Lensing effects can be observed over a wide range of scales: 
  from megaparsecs \cite[weak lensing,][]{Massey2007, Hoekstra2008, Okabe2010, Engelen2012, Mandelbaum2013, CFHT2013}, 
  to kilo parsecs \cite[strong lensing,][]{Treu2010,Suyu2010, Oguri2012, Coe2013, Newman2013, Kelly2015}, 
  down to parsec scales \citep[micro lensing][]{Muraki2011, Mao2012, Han2013, Gould2014}. 
It has been widely
applied in extragalactic astrophysics and cosmology, e.g., in
reconstructing the mass distributions of lenses
\citep{Mandelbaum2006, Umetsu2008, Oguri2012, Newman2013, Han2015},
detecting galaxies at high redshift
\citep{Richard2008, Jones2010, Jones2013, Stark2014}, measuring the
Hubble constant \citep{Paraficz2010, Suyu2013, Suyu2014, Liao2015}, amongst other applications.

Cosmological strong
lensing is an extreme manifestation of this process, in which
the mass density creating the potential---i.e., the lens, which is typically a
massive galaxy or a group or cluster of galaxies---is sufficient to
create multiple highly magnified and distorted images of background
sources.
The occurrence and morphological properties of these lensed
images reflect both the properties of the gravitational potential between
the source and the observer, and the lensing geometry.
It is a powerful probe of the central mass structure in galaxy
clusters and groups and offers unique constraints on such systems
\citep{Halkola2006, Sand2008, Newman2009, Limousin2010, Newman2011, Limousin2012, Bhattacharya2013, Grillo2015}.
Gravitational lensed arcs are used in a variety of cosmological
applications \citep{Kneib2011, Meneghetti2013}.  The frequency of
strongly lensed arcs on the sky reflects the abundance
\citep{Dalal2004, Li2006, Fedeli2007, Hilbert2007, Fedeli2010}, the
concentration \citep{Oguri2012, Sereno2013, Meneghetti2014} and
astrophysical properties~\citep{Peter2013, Rozo2008} of massive lenses, and the redshift distribution and properties of the
sources \citep{Wambsganss2004, Bayliss2011, Bayliss2012}.  Thus, arc
statistics help trace structure formation and can in principle
provide constraints on cosmological parameters.

However, to fully exploit the burgeoning samples of strong lenses now
being reported in various survey datasets
\citep{Bolton2006, Hennawi2008, Bayliss2011b, Bayliss2012, More2012, Hezaveh2013, Stark2013, Dye2014}
and expected from upcoming surveys \citep{Collett2015} progress is required 
on many fronts.   
For example, currently we are unable to definitely answer even simple questions
such as ``Is the observed number of giant arcs in large statistical samples consistent with theoretical expectations?". The mismatch between predictions and observations---the so-called ``arc-statistics problem''---has remained
unresolved for nearly twenty years
\citep{Bartelmann1998, Li2006, Meneghetti2008, Meneghetti2010, Meneghetti2013}.
Recently, Xu et al. (2016) have demonstrated that the lensing cross section
of a small mass-selected cluster sample, mostly at low redshift, is
consistent with theoretical expectations. This is encouraging, but also
distinct from answering the broader question of whether the arc statistics
of the entire cluster ensemble in the universe is well-reproduced by
simulations.
Long-standing issues, such as mass sheet degeneracy
\citep{Saha2000, Bradac2004, Scheinder2014}, also still lurk. 

In order to study the above topics systemically, we have developed a pipeline to produce realistic
strong-lensed images for cluster- and group-scale lensing systems. PICS is designed to bridge the gap between large-scale surveys for
strong lenses, and the largest cosmological N-body simulations now
run. The new generation of simulations---and those that will follow---offer, for the first time, both the volume necessary to
sample massive halos with good statistics, and the mass resolution to resolve the
profiles of typical cluster lenses; both of these features are necessary
for robust strong-lensing predictions. The pipeline includes a density
estimator based on a resampled Delaunay Tessellation
\citep{Schaap2000, Bradac2004b} technique, a deflection angles calculator based
on Fourier methods, a module to build images of sources to be lensed
and otherwise included in the light cone, a module for building the
images of lens member galaxies, and finally an open-ended module to
make the simulated images correspond to observations from a given
telescope or survey. PICS can in principle also be used to
simulate weak lensing by galaxy clusters, galaxy-galaxy lensing, and
substructure lensing near the main cluster or group lenses.  The
pipeline is optimized to produce large numbers of simulated images---
hundred of thousands, or even millions---in reasonable timeframes. This
is necessary to match the huge cosmological volumes probed by ongoing and future imaging
surveys that cover appreciable fractions of the entire
sky.

The statistics of strong lensing are subject to manifold nonlinear effects, many of which are best captured using fully
ray-traced N-body or hydro simulations. Examples include effects due
to the ellipticity of the lenses
\citep{Meneghetti2003b, Meneghetti2007}, baryonic matter in the lenses
\citep{Meneghetti2003a, Rasia2012, Killedar2012b}, lens substructure
\citep{Hennawi2007, Meneghetti2007}, structures on the line of sight
\citep{Wambsganss2004, Faure2009, Bayliss2014}, the mass spectrum
normalization \citep{Li2006, Fedeli2008, Fedeli2010}, and the properties
of dark matter \citep{Mahdi2014}, in addition to the properties of
sources \citep{Keeton2001a, Wambsganss2004, Gao2009}. Furthermore,
observational effects, of which the simplest is the correlated
distribution of image quality, transparency, and sky brightness across
multiple filters in a given survey dataset, will drive the
detectability of strong lensing features not just as a function of
brightness, but also lensed-image type, arc length-to-width ratio, lensed
source color, etc. Consequently, even understanding the selection
function of a strong lensing survey, let alone its scientific
application, is a challenging task. This challenge is best met by
passing realistically simulated images through the same selection
process as the real data, which, at least at optical wavelengths, is
typically some form of visual selection
\citep[e.g.,][]{Hennawi2008, Bayliss2011b, Wen2011, Stark2013, More2016}.

An auxiliary use for this lensing code has been to test the
preservation of morphological measurements, such as the Gini
coefficient \citep{Abraham2003}, under strong gravitational lensing.
In particular, as the new generation of space-based large survey
telescopes, e.g., WFIRST\footnote{http://wfirst.gsfc.nasa.gov}, come 
online, the number of observed strong lensing systems is expected to
expand into the thousands.  While such systems provide more detailed views
of the internal structure of galaxies at higher redshift than would
otherwise be possible, the challenge is in extracting useful morphological information from such data.
It will likely be necessary to develop morphological measurements that
are conserved under gravitational lensing and some elements of the
code described here have been used to test the the reliability of image
plane measurements of the Gini coefficient \citep{florian2015a}.
The literature is replete with software to simulate cosmological
strong lensing, much of which---in order to
focus on various physical and computational issues relevant to the
calculation---works in a theoretical framework that
does not make an explicit link to observations of lenses (e.g.,
\citealt{Killedar2012a},\citealt{Takahashi2011}). 
Several computational frameworks have been built around cosmologically
appropriate distributions of analytic halos, in order to predict
strong lensing statistics \citep{Oguri2009, Giocoli2012}. 
Several well developed software efforts are described in the current literature,
including GLAMER \citep{Metcalf2014} and
SkyLens\footnote{https://github.com/pmelchior/skylens}\citep{Meneghetti2008}. The latter is most
directly comparable to the algorithms presented here, and we use several of the
same basic approaches to the problem of simulating strong lensing as that code.
It is not our aim to argue for the efficacy or excellence of one package over
another; all such software efforts---including PICS---must make assumptions and
simplifications to enable computational efficiency.

Many early efforts
have understandably been forced to use simplified halo models, and
single source planes with analytic sources; such simplified systems do
not allow generation of simulated observations that appear authentic,
and so have in most part avoided attempting to generate such data. The
focus of this paper is to establish a strong lensing simulation
framework, which runs in the highly parellelized computer environment of our
collaboration, that is aimed at generating large samples of realistic mock
observations, primarily of group- and cluster-mass halos. The mass
distributions are drawn from extensive N-body simulations,
and the source data from the best available Hubble Space Telescope
observations, and analyses thereof. The capability for full many-plane lenses and sources
is enabled in the code, and significant effort has been expended in
order to `observe' the entire light cone simulations, producing image
data indistinguishable from real observations.

This paper is structured as follows: In Section~2, we briefly review the basic
lensing theory.  Details of how to produce lensed images
using the data from our cosmological simulations are presented in Section~3.  In Section~4, we focus on techniques for making the lensed images
realistic in properties and appearance.  An illustrative comparison
between our results and real observations is shown in Section 5.  We
discuss the implications and limitations of this image simulation
pipeline and conclude in Section 6.

\section{Basic Lensing Theory}
\label{sec:basic_lensing}

The full formalism described here can be found in 
\cite{Schneider1992} and \cite{Narayan1996}. 
Throughout the paper, the thin lens approximation is adopted. 
The dimensionless surface mass density of a thin lens plane can be written as
\begin{equation}
 \kappa(\vec{\theta}) = \Sigma(\vec{\theta})/\Sigma_{\rm crit},
\end{equation}
with the critical surface mass density $\Sigma_{\rm crit}=({c^2}/{4\pi G}) ({D_{\rm s}})/({D_{\rm d}D_{\rm ds}})$,
where $D_{\rm s}$ and $D_{\rm d}$ are the angular diameter distances from the source and lens to the observer respectively, 
$D_{\rm ds}$ is the angular diameter distance from the lens to the source, 
and $\Sigma(\vec{\theta})$ is the surface mass density of the lens. 
The lensing potential is given by
\begin{equation}
 \Psi(\vec{\theta}) = \frac{1}{\pi}\int {\rm d}^2\vec{\theta^{'}} \kappa(\vec{\theta^{'}}) {\rm ln}|\vec{\theta}-\vec{\theta^{'}}| \,\,.
 \label{eq:psi_kappa}
\end{equation}
The deflection angles are given by
\begin{equation}
 \vec{\alpha}(\vec{\theta})=\frac{1}{\pi} \int {\rm d}^2 \vec{\theta^{'}} \kappa(\vec{\theta^{'}}) \frac{\vec{\theta} -\vec{\theta^{'}}}{|\vec{\theta}-\vec{\theta^{'}}|^2}\,\,.
 \label{eq:alpha_kappa}
\end{equation}

Suppose we have obtained the surface mass density on a grid $\kappa_{ij} = \kappa[i,j]$, 
then the discrete version of Eq.~\ref{eq:alpha_kappa} is written as 
\begin{equation}
 \vec{\alpha_{ij}} = \frac{1}{\pi} \sum \kappa_{kl}\frac{\vec{x_{ij}}-\vec{x_{kl}}}{{|\vec{x_{ij}}-\vec{x_{kl}}}|^2}\,\,.
 \label{eq:alpha_kappa_ij}
\end{equation}
If the number of grid cells is large, the direct summation is unacceptably slow.  
Fourier techniques can be applied to improve the efficiency, because
the deflection angle can be written as a convolution of the convergence $\kappa(\vec{\theta})$ with a kernel
\begin{equation}
K=\frac{1}{\pi} \frac{\vec{x}}{|\vec{x}|^2}\,\,.
\label{eq:kernel_x}
\end{equation}
This allows the Fourier convolution theorem to be applied, 
hence
\begin{equation}
 \hat{\vec{\alpha}}(\vec{k})=\hat{\kappa}(\vec{k})\hat{\vec{K}}.
 \label{eq:alpha_fft}
\end{equation}
Then performing an inverse Fourier transform on the deflection angles in Fourier space, 
we obtain the deflection angles in real space. 
The complexity of a Fast Fourier transform (FFT) is only $N{\rm log}(N)$, 
which is much smaller than that of regular summation ($N^2$), therefore FFT
methods can speed up the computation of the deflection angle considerably. 
This approach is distinct from the tree-based methods of \cite{Meneghetti2010},
\cite{Rasia2012}, and \cite{Metcalf2014}, which offer overall greater
flexibility - for example in the handling of micro-lensing - at computational
cost. Here, the aim is to focus on cosmological strong lensing of extended sources across very large samples, for which we choose a more limited but efficient FFT method.

Using an FFT to determine the discrete
Fourier transform of the convergence $\hat{\kappa}(\vec{k})$ 
requires the convergence to be periodic on the lens plane, enforced by suitable application of boundary conditions. If the science goal is focused on large scales, periodic boundary conditions are applied.
If we want to investigate lensing on small scales, 
for example, strong lensing effects around the central regions of galaxy clusters, isolated boundary conditions are applied, following \cite{Hockney1988}.

Once the deflection angles at the lens planes are known, 
we can construct the lensing equation for different source planes. 
For example, in the case of a single lens plane and a single source plane, 
the lensing equation can be written as
\begin{equation}
 \vec{\beta}=\vec{\theta}-\vec{\alpha}(\vec{\theta})\,\,,
 \label{eq:lensing_eq}
\end{equation}
where $\vec{\theta}$ is the angular position of the lensed images on the image plane, and $\vec{\beta}$ is the angular position of the source images on the source plane.
Based on Eq.~\ref{eq:lensing_eq}, ray-tracing simulations can be performed from the observer 
to the lens plane and then to the source plane to produce lensed images. In the case of multiple lens planes, the lensing equation can be written as:
\begin{equation}
 \vec{\theta}^{S} = \vec{\theta}-\sum^{N}_{i=0} \frac{D_{\rm is}}{D_{\rm s}} \vec{\alpha}^{i}(\vec{\theta}^{i})\,\,,
 \label{eq:lensing_eq_mp}
\end{equation}
where $N$ is the number of lens planes, $D_{\rm s}$ is the angular
diameter distance from the observer to the source and $D_{\rm is}$ is
the angular diameter distance from the $i$th lens-plane to the source
plane. $\vec{\theta}^{S}$ is the angular position of the source on the
source plane, $\vec{\theta}$ is the angular position of the lensed image on
the image plane and $\vec{\alpha}^{i}$ is the reduced deflection angle on the
$i$th lens-plane.  Moreover, if there are also multiple (or many)
source-planes, which is typical for a full light cone simulation, one
needs to split such a lensing system into sub-systems each with a
fully ray-traced single source. These can then be stacked together to
produce the full lensed light cone.  Further discussion of multiple plane lensing systems can be
found in \cite{Hilbert2007} and \cite{Petkova2014}. Multiple lens planes and multiple source planes are
both enabled within PICS.


\section{Simulated Lensed Images}
\begin{figure}
	\centering
	\includegraphics[width=0.5\textwidth]{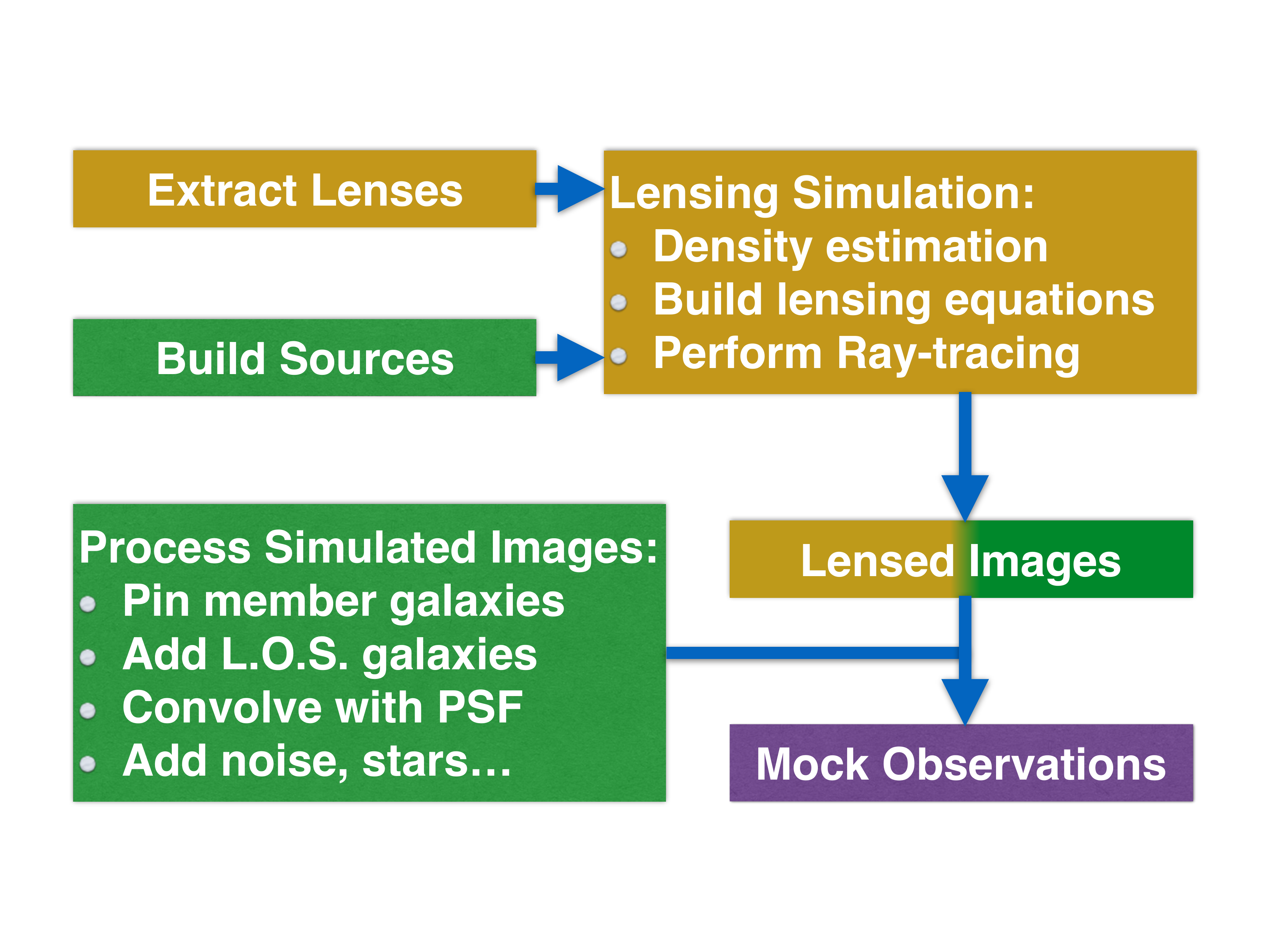}
	\caption{Flowchart of the strong-lensing pipeline - PICS. Yellow denotes simulation results, green, results based on real data, and the purple block is the final output. The simulated lensed images are a combination of real data from HST and the Outer Rim simulation (Cf. Section~\ref{sec:cosmosim}). The source light cone used real source images; simulated source images will be added in the future.
    }
	\label{fig:flowchart}
\end{figure}

Generally, simulations of strongly-lensed arcs include three major
steps. First, one requires a mass distribution for the lens or
lenses. These may be simple analytic models \citep{Keeton2001, Xu2009},
semi-analytic models \citep{Giocoli2012, McCully2014} or based on
particle distributions from numerical simulations
\citep{Dalal2004, Hilbert2007, Meneghetti2008, Meneghetti2010, Rasia2012, Metcalf2014}. 
PICS is designed to work in the latter regime, with
N-body simulations providing mass distributions. As discussed above,
one of our principal motivations is arc statistics in large surveys,
and simulations naturally capture many of the effects which drive arc
counts and properties.  With a particle distribution, one also
requires a density estimator to calculate the surface mass density map
of each lens plane.  With a mass map in hand, the next major step is
to calculate the deflection field according to the normalized surface
density and then build the lensing equation.  Finally, ray-tracing
is performed to map the locations of pixels in the image plane to the
source plane, and---after calculating the flux information of each traced
light ray (i.e., by placing images of sources in the source plane)---
light rays are mapped from the source plane back onto the lens plane 
to complete the lensed image. In Figure~\ref{fig:flowchart} we provide a flow chart overview of our PICS pipeline and describe each step in detail below. 

\label{sec:lensed_images}
\subsection{Cosmological Simulations}
\label{sec:cosmosim}

The cosmological simulation results used in this paper have been
obtained with the Hardware/Hybrid Accelerated Cosmology
Code \citep[HACC,][]{habib14}, a flexible, high-performance N-body code that
runs on a range of supercomputing architectures. In this case, we used
Mira, a BG/Q system at the Argonne Leadership Computing Facility to
carry out the ``Outer Rim'' simulation, one of the largest cosmological simulations
currently available.

The cosmology used is a $\Lambda$CDM model close to the best-fit model
from WMAP-7~\citep{wmap7}. In detail, the cosmological parameters are:
$\omega_{\rm cdm}=0.1109$, $\omega_{\rm b}=0.02258$, $n_s=0.963$,
$h=0.71$, and $\sigma_8=0.8$. The box size of the simulation is
$L=4225.4~{\rm Mpc}=3000~h^{-1}{\rm Mpc}$, and it evolves 10,240$^3$
=1.07 trillion particles. This leads to a particle mass of
$m_p=2.6\cdot 10^{9}~{\rm M}_\odot=1.85\cdot 10^9~h^{-1}{\rm
  M}_\odot$. Extensive testing (Rangel et al. 2015)
indicates that at this mass resolution we are able to robustly compute
strong lensing for halos of masses $M_{500c}> 2\cdot
10^{14}h^{-1}$M$_\odot$ (where $500c$ denotes the overdensity relative to the
critical density). The simulations naturally incorporate substructure
to much smaller mass scales; at these small
scales the effects of (primarily stellar) baryons will play a role in the
formation of small-scale lensing features. Future comparisons between the
outputs of this pipeline and real data will consider such effects.

The large volume ensures that we have high-mass clusters
at early times present in the simulation, and extensive statistics for
massive systems which individually have large lensing cross sections. 
Unlike \cite{Meneghetti2008, Meneghetti2010} the baseline simulation used here is sufficient to enable strong lensing calculations without any re-simulation.
For demonstration purposes to illustrate such systems in the latter
sections of this paper, we have chosen three massive cluster halos at
three different redshifts ($z=0.736,~0.539,~0.364$) with masses representative of observed South Pole Telescope (SPT) strong lenses \citep{Bleem2015}. Further discussion can be found in Section~\ref{sec:results}.

Halos are identified with a Friends-of-Friends (FOF) halo finder with a linking length of $b=0.168$, versus the canonical value of $b=0.2$, following~\cite{cohnwhite} who found that this reduced value mitigates the problem
of halo overlinking. Overdensity masses $M_{500c}$ of the
clusters in the simulation are computed based on the centers of the
FOF halos. These centers are determined by finding the potential
minimum for each cluster.

We have identified the subhalos in each cluster using a
phase-space based subhalo finder.  There are a number of ways to identify substructure in halos; two commonly used ways are running a hierarchical FOF algorithm \citep[as used in, e.g.,
Rockstar,][]{rockstar} or using a local overdensity finder \citep[as
used in, e.g., SUBFIND,][]{subfind}. Such approaches can be
enhanced by also using velocity information to determine if a particle
actually belongs to a subhalo or is a member of the main halo (e.g., the
subfinding algorithm in Rockstar). Our approach is based on local density
estimation and folds in phase-space information as well. We first
organize all particles that were found by the FOF halo finder into a
Barnes-Hut tree structure~\citep{barneshut} to provide them in an
easily accessible format.  A smooth particle hydrodynamics (SPH)
kernel is then used to estimate the local density for each particle
(very similar to the original SUBFIND approach).  Particles are
assigned into subhalo candidates and we build an initial membership
list for the main halo as well as subhalo candidates. The final step
is to determine if a particle is bound to a subhalo or in fact belongs
to the main halo. Here we calculate the total energy for each particle (kinetic
plus potential) to determine its escape velocity with
regard to the subhalo. We investigate each particle
one by one, testing if it belongs to the subhalo, and if it does not,
we assign it to the main halo. If a particle gets removed from the
subhalo, a new energy calculation is carried out and the remaining
particles are investigated. This procedure continues iteratively
until only bound particles remain in the subhalo candidate. Finally,
we set a minimum number of particles for a subhalo to be considered
viable (20 particles in this case) and subhalo candidates with fewer
particles than this limit are discarded. 
Each final subhalo catalog provides positions, orientations,
ellipticities and masses, and this data is used in combination with
data from real observations to ``paint'' cluster member
galaxies onto simulated images (Section 4). 

\subsection{Density Estimator}
Our algorithm for computing the surface density from a halo particle set 
implements the Delaunay tessellation field estimator (DTFE) method proposed by 
Schaap and Weygaert~\citep{Schaap2000}. The DTFE method requires the Delaunay triangulation\footnote{A Delaunay triangulation of a discrete 3D point set
  $\mathit{P}$ is a triangulation $\mathit{DT \left ( P \right ) }$
  such that the circumscribing sphere of a tetrahedra contains no
  other points in $\mathit{P}$. } of the halo particles to obtain the
local gradients of the 3D density function.  This first order
interpolation method assumes the gradient within a tetrahedron to be
constant and discontinuous at its boundaries. The field value for an
arbitrary point, $\vec{x}$, is interpolated using the Delaunay
vertices $\vec{x}_{0} , \vec{x}_{1}, \vec{x}_{2}, \vec{x}_{3}$ of the
containing tetrahedron by
\begin{equation} \label{interpolation}
\widehat{f(\vec{x})} = f(\vec{x}_{0}) + \widehat{\nabla f}
\rvert _{Del} \cdot (\vec{x}-\vec{x}_{0}),
\end{equation}
where $\widehat{\nabla f} \rvert _{Del}$ is the estimated constant
field gradient using the known field values
$\mathit{f(}\vec{x}_{0}), \mathit{f(}\vec{x}_{1}), \mathit{f(}\vec{x}_{2}), \mathit{f(}\vec{x}_{3})$. For density field reconstruction,
the on-site density values are estimated by the inverse volume of the
contiguous Voronoi cell, thereby ensuring mass conservation. The
estimated density for each input, $\vec{x}_{i}$, is given by
\begin{equation}
	\widehat{\rho(\vec{x}_{i})} = {{(d+1) m} \over
          \sum_{j=1}^{N_{\mathcal{T},i}} V({\mathcal{T}_{j,i}})},
\end{equation}
where $N_{\mathcal{T},i}$ denotes the number of tetrahedra
$\mathcal{T}_{j,i}$ having $\vec{x}_{i}$ as a vertex, $m$ is the
mass, and $(d+1)$ is the tetrahedral volume normalization factor.
 
We compute the surface density field as a 2D uniform grid, where each grid cell 
represents the average surface density of vertical paths filling a $\Delta x \times \Delta y \times \ell$ 
vertical column in the 3D density field. 
From the equation for surface density along a vertical path, given by
\begin{equation} \label{eq:vpath}
	\Sigma(\vec{\xi}) = \int  {\rho(\vec{\xi}, z)\ dz},
\end{equation}

\noindent where $z$ is the vertical coordinate, and $\vec{\xi}$ is a 2D field point, it follows that the surface density value for a grid cell covering the region $S$ is given by

\begin{equation} \label{eq:surfing}
	\overline{\Sigma}_{S}  = {{\int_{\Delta y} \int_{\Delta x} \Sigma(\vec{\xi})\ d\xi_1 \ d\xi_2} 
	\over {\int_{\Delta y} \int_{\Delta x} d\xi_1 \ d\xi_2 }}, \ \vec{\xi} \in S.
\end{equation}

In our algorithm, we calculate the optimal DTFE value for Eq.~\ref{eq:vpath} by 

\begin{equation}
	\widehat{\Sigma(\vec{\xi})} = \sum_{i=1}^{N_{\mathcal{T},\vec{\xi}}}
         \widehat{\rho(\vec{\xi},mid(\mathcal{T}_{i,\vec{\xi}},\vec{\xi}))} \  len(\mathcal{T}_{i,\vec{\xi}},\vec{\xi}),
\end{equation}

\noindent where $N_{\mathcal{T},\vec{\xi}}$ denotes the number of tetrahedra $\mathcal{T}_{i,\vec{\xi}}$
intersecting the vertical path of $\vec{\xi}$, $mid$ is a function for determining the midpoint of a line-tetrahedron intersection, and $len$ is a function for determining the length of the intersection. We use the optimal DTFE surface density in a Monte Carlo (MC) approximation for Eq.~\ref{eq:surfing}, where the number
of MC samples is weighted by the number density of input particles in the 3D vertical column.

\subsection{Making A Catalog of Source Images}

Source galaxies are selected from the Hubble Ultra Deep Field
\cite[HUDF]{Beckwith2006} informed by the \cite{Coe2006} photometric
redshift analysis, similar to elements of \cite{Meneghetti2010} and \cite{Rasia2012}.
The HUDF data offer high spatial resolution
relative to the resolution of the ground-based telescopes that our
image simulations are intended to imitate and are sufficiently deep
that they sample the high-redshift galaxies which form the bulk of the
strongly-lensed population. 

The PICS pipeline is focused on producing
mock observations for typical ground-based telescope imaging, for which the
relevant HST-imaged galaxies are reasonably bright (and large). Hence, unlike
\cite{Meneghetti2008, Meneghetti2010} and \cite{Rasia2012} we do not attempt to
deconvolve the Hubble images or treat the intrinsic noise of those images; the
pipeline as presented here uses the HUDF images directly (see the
discussion in Section 6).
One further drawback of the dataset is that the sampled area
is relatively small, and so cosmic variance is a possible
concern. This is also discussed further in Section~6 below.	

In the simulated images shown below, we have limited our analysis to
three HUDF bands: F435W, F505W, and F775W. Individual galaxies in the
HUDF were selected by first stacking the images in these three bands,
and then cutting at a threshold of 2.0$\sigma$ above the background
noise level and keeping only pixels that had at least three
neighboring pixels that were also at or above the threshold.
Any pixels that were removed by this initial cut, but that were
surrounded on all sides by pixels that were not removed and that all
belonged to the same object were also kept. The final set of pixels
for each source, which we refer to as an {\it aperture mask}, then defines
`live' pixels which must be considered when ray-tracing a particular
source. The complete list of potential sources was then refined to
include only those that match objects in the photometric redshift
catalog of \cite{Coe2006}. Our source catalog covers about $78\%$ of the objects in Coe's catalog. That analysis is a detailed look at the
HUDF data, including redder bands not considered here, and is more
than sufficiently deep to sample even the faintest sources relevant
for simulating ground-based survey imaging. 

Once an aperture mask is made for all viable source galaxies,
subfields of the HUDF are chosen to become part of the light cone for
gravitational lensing in the ray-tracing code. The width of these
subfields was chosen to be about four times larger than the largest
caustic structures in the simulation, resulting in subfields of
$1024\times1024$ pixels. The centers of subfields are chosen randomly,
but with the requirement that the entire $1024\times1024$ pixel region
contains data from the HUDF (i.e., does not extend past the edges of
the field).

The redshift of each galaxy in a given subfield is taken to be the
maximum-likelihood value from the photometric redshift catalogue of
\cite{Coe2006}. To streamline computation, source galaxies are binned
by redshift. Because the effects of gravitational lensing (e.g., the
size of the Einstein radius) vary most rapidly at redshifts that also
happen to have the highest galaxy counts, the bins were selected using
a simple calculation such that each bin has an equal number of
galaxies, which is set as ten in default, 
except for the highest redshift bin, 
which was allowed to have fewer galaxies than the others. 
This allows for narrow redshift
bins when lensing effects are rapidly changing, but also minimizes the
number of redshifts for which deflection angle maps were required. A
complete ray-trace is then performed for the median redshift of each
bin, rather than for each source. All galaxies in each redshift bin 
are treated as if they were at that bin's median redshift.

\subsection{Performing Ray-tracing Simulations}
\label{sec:performing}
\begin{figure}
	\centering
	\includegraphics[width=0.475\textwidth]{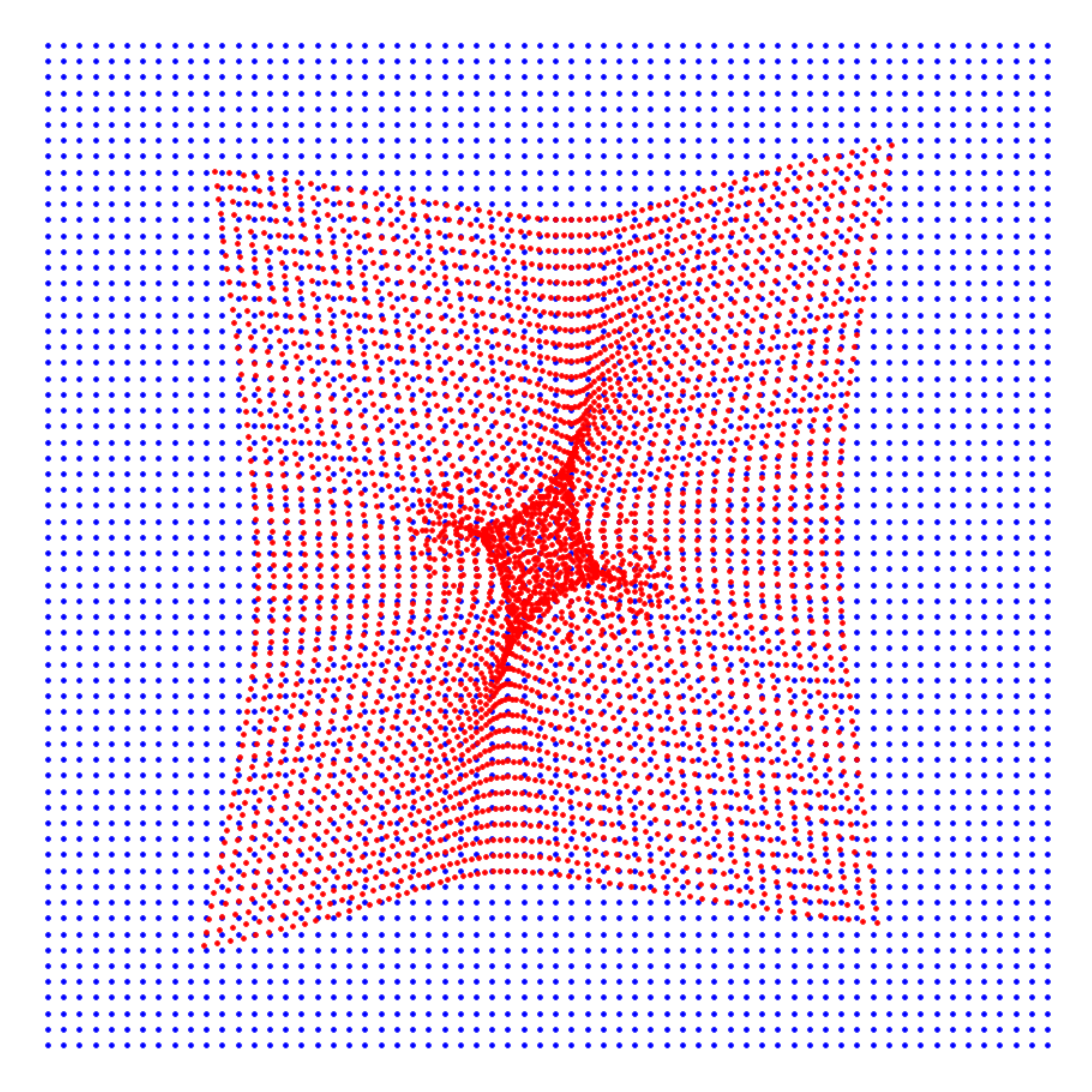}
	\caption{Illustration of mapping the lens plane to the source plane.  Blue points show the positions of the intersection between light rays which start from the observer, and the image plane.  To produce the lensed images using Fourier methods, we sample the blue points on regular grids.  Red points illustrate the intersection between the deflected light rays and the source plane. 
}
	\label{fig:mapping}
\end{figure}
With the surface mass density of the lens or lenses in hand, as well
as a set of source planes, the next step is to compute the deflection
angle field at each lens plane. To do this, we first build a grid at
each lens plane. Because we
use Fourier methods to boost the efficiency of the convolution (see 
Eq.~\ref{eq:alpha_kappa_ij}), we use a regular mesh\footnote{The use of adaptive meshes is also an alternative option for increasing computational speed.  However, with a
static grid and Fourier methods, we prefer to simply decrease the pixel
size on the lens plane to make higher quality simulated images if and
as necessary, since computational cost is not a limiting factor
compared to the resources necessary to make the input simulations.}.
For each lens plane on the grid we then compute the deflection angle
field at each position. 

When light rays cross the lens plane, they are deflected
with an angle
\begin{equation}
\tilde{\vec{\alpha}} = \frac{4G}{c^2} \int {\rm d}^2\vec{\xi^{'}} \Sigma(\vec{\xi^{'}})\frac{\vec{\xi}-\vec{\xi^{'}}}{|\vec{\xi}-\vec{\xi^{'}}|^2}\,\,,
\label{eq:alpha_hat}
\end{equation}
where $\Sigma(\vec{\xi})$ is the surface mass density on the lens plane,
$\vec{\xi}$ is the physical position on the lens plane, and 
$\tilde{\vec{\alpha}}$ is the physical deflection angle---i.e., it is the
angle between the extension line of incoming light rays and the
outgoing light rays at the lens plane.  The relation between the
physical deflection angle $\tilde{\vec{\alpha}}(\vec{\xi})$ and the reduced
deflection angle $\vec{\alpha}(\vec{\theta})$ is written as
\begin{equation}
 \vec{\alpha}(\vec{\theta}) = \frac{D_{\rm ds}}{D_{\rm s}}\tilde{\vec{\alpha}}(D_{\rm d}\vec{\theta})\,\,.
 \label{eq:alpha_scale}
\end{equation}
The deflected light rays travel to the source plane, and intersect
with the source plane.  For the single image plane and single source
plane case, we can use Eq.~\ref{eq:lensing_eq} to find the source
plane positions directly. For multiple lens plane and multiple source
plane cases, the physical deflection angle is calculated first, and we
then use the relation in Eq.~\ref{eq:alpha_scale} to scale the
physical deflection angles to reduced deflection angles, and the
positions in the source plane can be determined by using
Eq.~\ref{eq:lensing_eq_mp}. We are now ready to ray-trace and
make lensed images on the image plane\footnote{Note that in a system
  with a single lens plane, the ``image'' plane and the ``lens'' plane
  coincide. This is not generally true; the image plane is
  coincident with only the lowest redshift lens plane in more complex
  systems.}.

On the image plane, the pipeline default output pixel size and scale for output
ray-traced images is 0.09~arcsec/pixel, and 2048$\times$2048 pixels. This
output sampling is sensibly coarser than the input source plane
sampling, coarse enough to provide a reasonable field of view with
2048$\times$2048 pixels (just beyond 3$\times$3 arcminutes, sufficient
for even very massive lenses) while still providing a much finer
sampling than the typical ground-based telescope point spread
function. 

With the image plane established, the second step is to use the deflection
angle fields established above to trace light
rays from the image plane to the source plane.  Figure~\ref{fig:mapping} shows an
illustration of tracing light rays from the observer through the image
plane to the source plane in the simple case of a single lens plane.
We assume that the grid points (blue points) are the intersections of
the image plane and the light rays which start from the observer.

\begin{figure}
	\centering
	\includegraphics[width=0.475\textwidth]{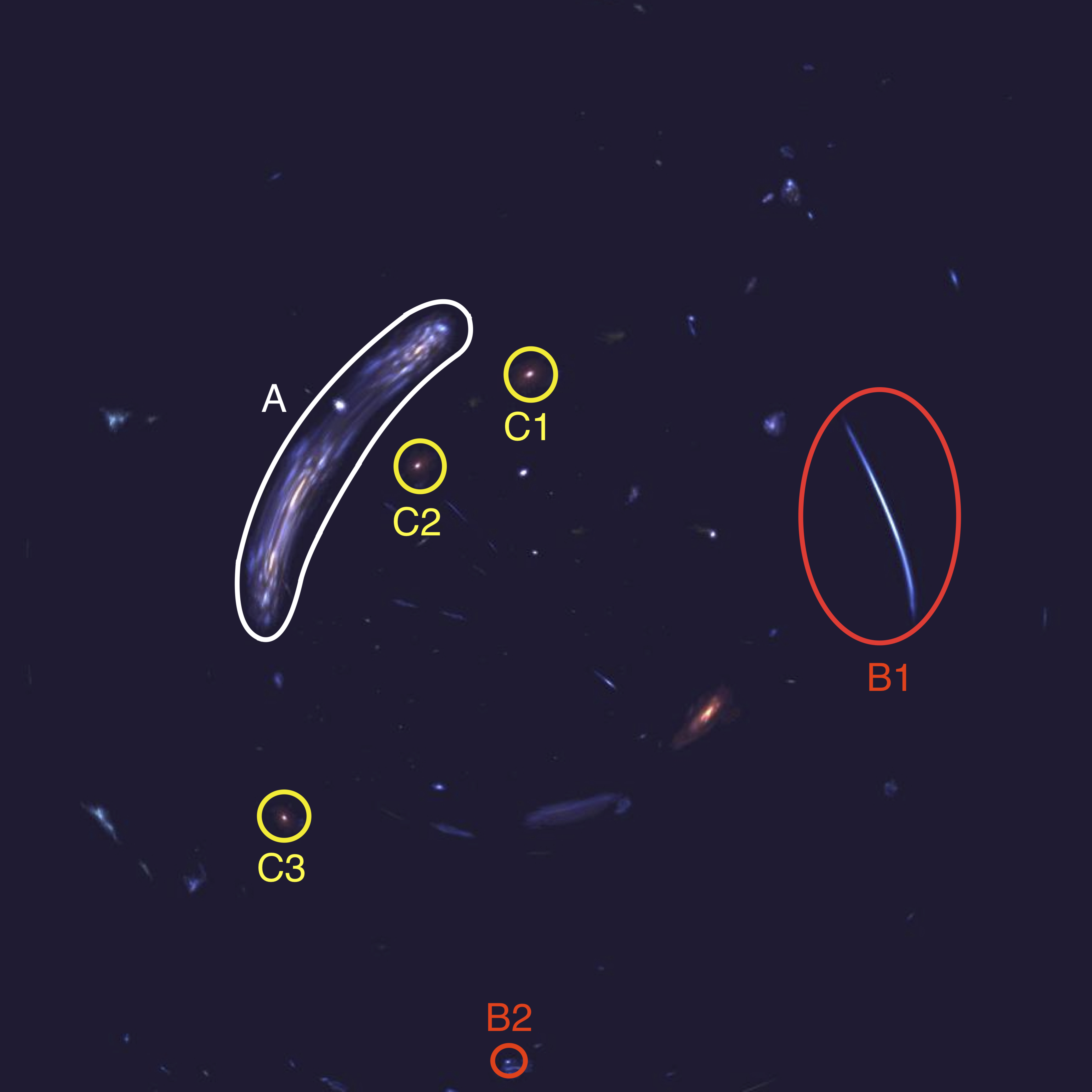}
	\caption{An example simulated lensed image with $900\times900$
          pixels at a sampling of 0.09~arcsec/pixel. The lensing is
          due to an $M_{500}=4.5\times10^{14} M_{\odot}$ halo at
          z=0.539.
The white aperture (A) marks a lensed giant arc, with the source in this
case located near a naked caustic cusp; the red apertures (B) mark a merging pair image forming a giant
lensed arc and a third image of the same source, with the source of these images
located around a fold caustic; the yellow circles (C) mark three
lensed images, of which the source is located near the same naked
caustic cusp as the source in image A.}
	\label{fig:lensed_imgs}
\end{figure}

With the lensing geometry now fully described for each pixel of the
image plane, the final step is to propagate brightness information
from the source plane back to the image plane. The brightness of each
ray is computed from the source plane brightness information as
described by the cataloged HUDF source images. Since the rays
intersecting the source plane sample the pixelated source plane
irregularly,
interpolation is applied.  Three interpolation schemes have been
considered: linear interpolation, bicubic interpolation and
bicubic-spline interpolation. We conclude that the improvement
from either bicubic and bicubic-spline interpolation is limited, and---as we are
focused on arc statistics from ground-based imaging data---not 
worth the extra computational time. Linear
interpolation is thus the default method.
After calculating the brightness for each traced light ray, that data
is mapped back to the image plane. Following Figure~\ref{fig:mapping}, we now have the brightness at each red point, the one-to-one
relationship between the blue points and red points, and hence we
obtain the brightness at the blue points, which is simply the
pixelated image of the image plane. For the typical case of multiple
source planes, sampling the redshift distribution of sources, this
entire process is iterated across the redshift bins and the resulting
image sub-planes stacked to produce the final result. An example simulated lensed image is shown in
Figure~\ref{fig:lensed_imgs}. In this simulated lensing system, an
input light cone of sources is picked from the HUDF as described above
with an area of $30\times30$ arcseconds. The surface mass density of
the lens plane is calculated by applying our improved density
estimator to a cluster-sized halo from the Outer Rim simulation.
The $M_{500c}$ of this halo is $4.8\times10^{14} M_{\odot}$, and
its redshift is 0.539.  For clarity, in Figure~\ref{fig:lensed_imgs} we show only
the central $81\times81$ arcseconds or $900\times900$ pixels at a
sampling of 0.09 arcsec/pixel; the computed image from which this is
drawn is $2048\times2048$ pixels, as described above.

\section{Adding the Lightcone and Cluster Galaxies}
\label{sec:decoration}

The previous sections describe in detail how the image simulation
pipeline produces only strongly lensed images. To make simulated images
that appear realistic, we must further add several other image
components and simulate observation by a telescope and camera
system. Elements of this portion of the overall simulation pipeline
include
\begin{itemize}

\item galaxies belonging to the lens itself (which, in the case of a massive and/or low-redshift cluster lens and
  shallow observations, can dominate the
  image),

\item other galaxies along the line of sight which are not
  significantly lensed (because they are either in front of the lens
  plane, or behind it but at large impact angles) and

\item faint stars.

\end{itemize}

 The final image stack is then passed through an open-ended module
 which ``observes'' the field; basically this process entails: convolution with a
 telescope point spread function (PSF) model, re-binning to the final
 desired sampling, and adding noise elements that replicate the real
 data being mimicked (including elements such as chip defects and
 cosmic rays, if required). Bright stars, which typically show
 telescope- and camera-specific oddities such as chip bleeding, are
 added at this final stage and treated separately, as discussed in
 below in Section~\ref{sec:brightstar}.

\subsection{Lens Member Galaxies}
The principal focus of our work is lensing by galaxy clusters and
groups, and what follows is appropriate to ``painting'' galaxies on to
dark matter halos of that mass scale. While the strong lensing
pipeline can in principle be used to describe lenses
of arbitrary masses, at single-galaxy mass scales inputs other than
large scale cosmological N-body simulations would be appropriate for
generating the strong lensing signals, and we do not attempt to model such systems here.

As it is our intention to use PICS to model and
analyse extensive survey data, the fundamental methodology we have in
mind is to draw photometric information about cluster and group member
galaxies from such survey data. However, such a treatment is left for
future papers, as it will be perforce a survey- and data-specific
task. To make progress here, and demonstrate the utility of PICS, we instead draw photometric information from the Second
Red-Sequence Cluster Survey \citep[RCS2]{Gilbank2011} as inputs into
simulated cluster-scale lensing images. Specifically, we use the
background corrected photometry of several massive clusters in the
RCS2 at $0.5<z<0.6$, the richness of which indicate masses
comparable to galaxy clusters in the South Pole Telescope Survey
cluster sample of \cite{Bleem2015}, as the fiducial values for cluster
member magnitudes in the $griz$ bands (the native observed bands in
the RCS2 data). The cluster member photometry is corrected, to first
order, for measurement scatter in the fainter galaxies by moving
galaxies toward the red-sequence colors that dominate cluster members
in cluster cores. We do this by keeping the $r$-band magnitudes fixed
and perturbing both the $g$- and $i$-bands toward the red sequence
colors (i.e., redder galaxies become bluer, bluer galaxies conversely
become redder). Individual magnitudes are changed by a one-sided
Gaussian random draw from the measured magnitude errors, hence with
the presumption made that most galaxies in truth have colors closer to
the red sequence, and have scattered away from it in the input RCS2
catalog due to measurement uncertainties. We also clean the input
photometric catalogs of objects with extreme colors---these are
overwhelmingly faint objects that are poorly measured to be extremely
red or blue in $g-r$, $r-i$, or $i-z$. Section~\ref{sec:results}
demonstrates simulated images of strong lensing from $0.3<z<0.8$. For
clusters at redshifts different than the input photometry, a simple
band-dependent shift in magnitudes is applied to make the magnitudes and colors of galaxies
on the red-sequence correct for the simulated redshift.
These are derived with $k$-corrections based on an elliptical galaxy spectrum in the same cosmology as the Outer Rim simulation. Bluer cluster
members will not be exactly correct, but for demonstration purposes
this is sufficient.

The photometric catalog is then matched to the subhalo catalog drawn
from the N-body simulation, with assignments made at
random. Typically, the list of observed cluster members is longer than the
list of subhalos, since the input cluster photometry probes to masses
smaller than the subhalo catalog; the subhalo list is padded to the
appropriate length using a random draw of particle positions from the
main halo.

An algorithm that matches the cluster member
photometry to the subhalos by rank ordering, with the most luminous
galaxy assigned to the most massive subhalo, and so on, will suffer from bias due to the dynamical erosion of subhalo
masses as they are accreted into cluster cores, an effect which will
preferentially decrease the mass-to-light ratio of cluster members
near the cluster center. In the future, we will use subhalo mass and
photometry matching via subhalo-finding algorithms
that track subhalos temporally through the entire multi-petabyte
outputs of the Outer Rim simulation, allowing
the spatially dense luminous component to be assigned to dark matter
subhalos based on their masses upon accretion into the larger system,
before significant mass-loss occurs. Note also that the brightest
cluster galaxy is assigned to the center (potential minimum) of the main halo.

Before simulating images of cluster galaxies, we must further specify
the sizes, shapes, luminosity profiles and orientations of the cluster
galaxies. The orientations and ellipticities are taken from the
subhalo catalog. Those galaxies assigned to random halo points are
given an arbitrary orientation; ellipticities are drawn at
random from a distribution shaped to approximate the distribution of
projected axis ratios for elliptical galaxies as found by
\cite{Ryden1992}. 

The radial light profiles of all galaxies are described by S\'ersic
profiles---a S\'ersic index of 1 for pure disks, and ranging from 3 to
6.2\footnote{The upper limit of 6.2 is adopted due to limitations of the image simulation package used to produce member galaxies - see below.} for the bulge component---as well as a bulge-to-total ratio (B/T)
for each galaxy. All galaxies with colors in $r-z$ redder than 0.2
magnitudes bluer than the red-sequence are assigned a bulge-to-total
ratio in $z$-band of $B/T_z=1$. Galaxies with $r-z$ bluer than
$r-z$=0, effectively the blue edge of the cluster galaxy distribution,
are assigned a bulge-to-total ratio in the $z$-band of
$B/T_z=0.1$. Galaxies in between these two limits are assigned
$z$-band bulge-to-total ratios linearly interpolated between 0.1 to 1,
according to their $r-z$ color. The bulge-to-total ratios in the bluer
filters are then assigned as a power of the $z$-band value: for
example the most different is the bluest filter,
$B/T_g=B/T_z^{1.25}$. This ensures that bluer galaxies observed in the
bluer bands are more `disky' than in the redder bands, while the pure
bulge---i.e., elliptical---galaxies are such in all four filters. The
bulge component of galaxies for galaxies fainter than $M_z^*+1$ are
assigned a S\'ersic index of 3, and the brightest cluster galaxy is
assigned a S\'ersic index of 6.2, with values interpolated between these
limits according to the $z$-band magnitude, consistent with the
distribution observed in low redshift galaxy cluster members
\citep{Fasano2012}.

The sizes of the bulge component of cluster member galaxies with flux
$F$ are assigned simply as $C+1/2\log{F}$, with the constant C chosen
such that the effective radius of the largest galaxies is $\sim$8~kpc,
consistent, for example, with that seen for the local giant elliptical
M87 \citep{Murphy2011}. Disk sizes, similarly, are assigned using a
linear relationship between galaxy area and luminosity to replicate
the disk size distribution seen in \cite{Laurikianen2010}.  The
brightest cluster member is allowed to vary up to one magnitude brighter or
fainter than the RCS2 fiducial, and is described by a two component
S\'ersic model where one component has a size 5$\times$ that of the
other, approximately simulating the extended stellar halo often seen
around central galaxies. The luminosity fraction assigned to this
component is allowed to vary from 25-75\% of the total galaxy light.

The details of the size and light profile assignments sketched above
are admittedly ad hoc. However in simulating optical images of
intermediate redshift cluster galaxies, observed from the ground, they
are sufficient. Actually measuring the morphology of such galaxies
requires Hubble Space Telescope observations \citep{Dressler1997} and
so in truth the exact details---at the level of the fundamental plane
for example \citep[e.g.]{Kormendy2009}---matter little in creating
images that look realistic in this context. Additionally, we remind
the reader that the intended use of this image simulation pipeline is
primarily to create large sets of simulated images matched to large
sets of actual survey data, from which the distributions of properties
of the lens galaxies would be directly drawn; the simulation approach
sketched above serves only to allow illustration of the pipeline
absent this close coupling to a particular survey dataset.

\begin{figure}
	\centering
    \includegraphics[width=0.475\textwidth]{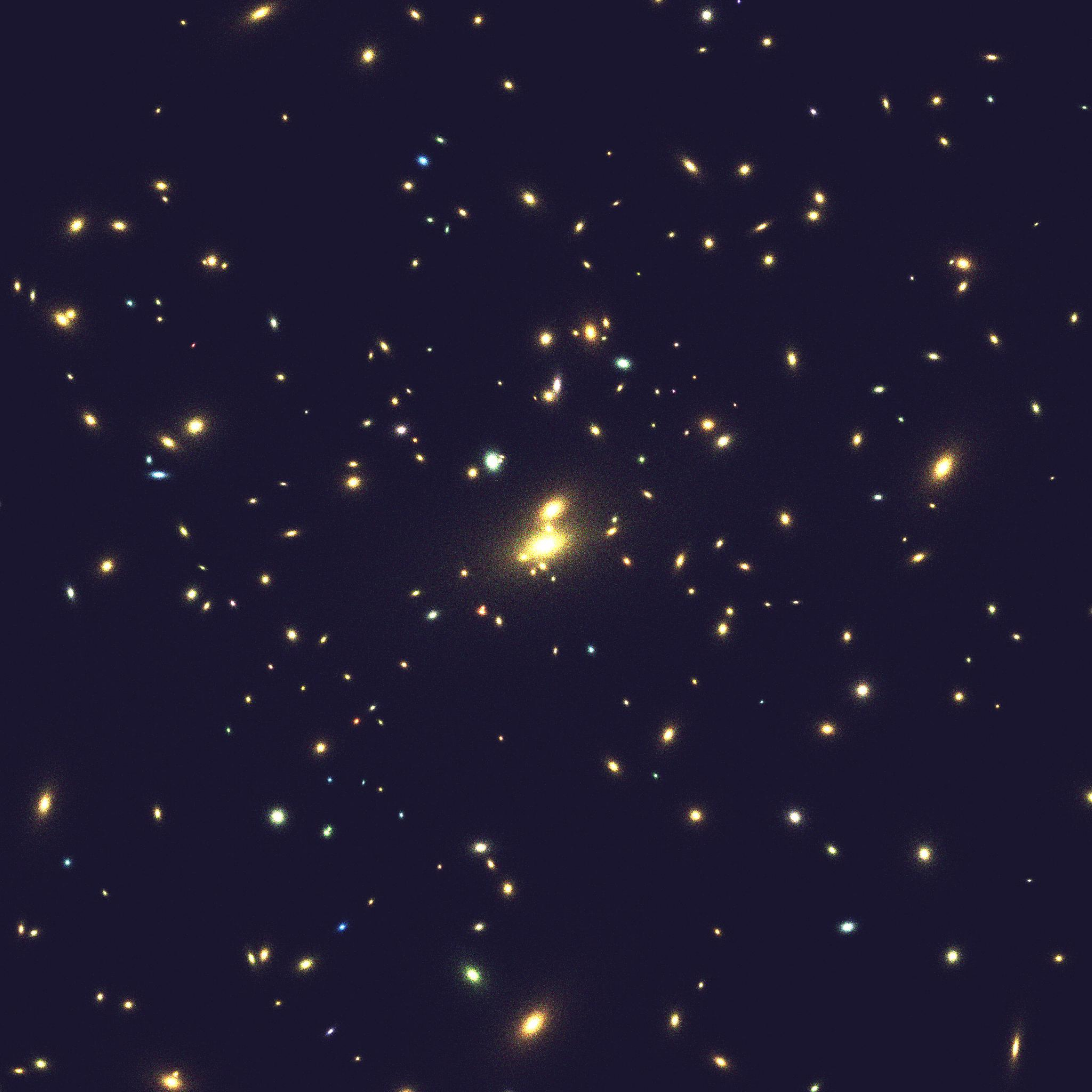}
	\caption{An example of simulated member galaxies of a cluster
          lens.  The size of this field is approximately 3'$\times$3',
          i.e. with $2048^2$ pixels at an angular size per pixel of
          0''.09.}
	\label{fig:mem_imgs}
\end{figure}

With a list of observable properties in-hand (the sources of which
are summarized in Table~\ref{tab:galsim}) the actual images of lens
member galaxies are made using the open-source {\tt GalSim} package
\citep{Rowe2014}, an astronomical image simulation toolkit designed
primarily to enable characterization of weak-lensing estimators with
accuracy sufficient to meet the requirements of future Stage IV Dark
Energy surveys. Here we simply use the mock galaxy generation routines
to create images of the member galaxies following the
parameterizations sketched above.  We use the “photon shooting”
method, in which the surface brightness profile of the galaxies are
finitely sampled, and generate a high-resolution,
low-noise\footnote{The Poisson noise inherent in the photon-shooting
  step is subdominant to the observational noise added
  below.} fits image of the cluster in each of our desired filter
bands.  An example of the output from the {\tt GalSim} routine is
shown in Figure~\ref{fig:mem_imgs}.

\begin{table}[h]
\centering
\resizebox{0.475\textwidth}{!}{
\begin{minipage}{0.475\textwidth}
\centering
\begin{tabular}{| c | c |} 
\hline 
Property & Source\\ \hline \hline 
Position & Outer Rim Simulation \\ \hline
Orientation & Outer Rim Simulation\\ \hline 
Axis Ratio & Outer Rim Simulation \\ \hline 
Multi-filter magnitudes & RCS2 Data \\ \hline
Bulge to disk ratio (per filter) & ad hoc \\ \hline
Bulge half light radius & Fit to Literature \\ \hline 
Disk scale radius & Fit to literature \\ \hline 
Sersic index & Literature \\ \hline \hline
\end{tabular}\par
\bigskip
\caption[{\tt GalSim} input sources.]{Properties of Lens Member Galaxies and Provenance.} 
\label{tab:galsim} 
\end{minipage}}
\end{table}

\subsection{Adding Galaxies on the Line of Sight}

Galaxies on the line of sight that are not strongly lensed are another
important component of realistic strong lensing images.
One method to add these is to build a full light cone with galaxies on
the line of sight, and then ray-trace as above through the whole light
cone.  The advantage of this method is that the lensed images that
result reflect all parts of the lensing signal (i.e., both strong
and weak lensing) and that the results are accurate, especially in
the weak-lensing regime, where such techniques are widely applied
\citep{Jain2000, Vale2003, Hilbert2009, Meneghetti2008, Meneghetti2010, Rasia2012, Killedar2012a, Metcalf2014}. The disadvantage
of this method is its computational cost.

The other method, and the one used here, is to fully ray-trace only a
small light cone which covers the caustics of the lenses, where strong
lensing happens. This provides images of any galaxies that are
significantly magnified, but drops weak lensing signals far from the
center of the lens.  The next step is to build a light cone image
without ray-tracing, and finally to stack both together. The advantage of
this method is its computational efficiency; this is especially important 
for studies of strong lensing effects since---as strong lensing is a rare phenomenon---many 
instances must be run. This is the basic method adopted here. 
However, it should be noted that the PICS code is flexible, in that the size of
the region ray-traced is arbitrary, and can be set to match the requirements of
the problem being addressed. A weak-lensing application, for example, would use
a much larger ray-traced light cone.

\begin{figure}
	\centering
	\includegraphics[width=0.475\textwidth]{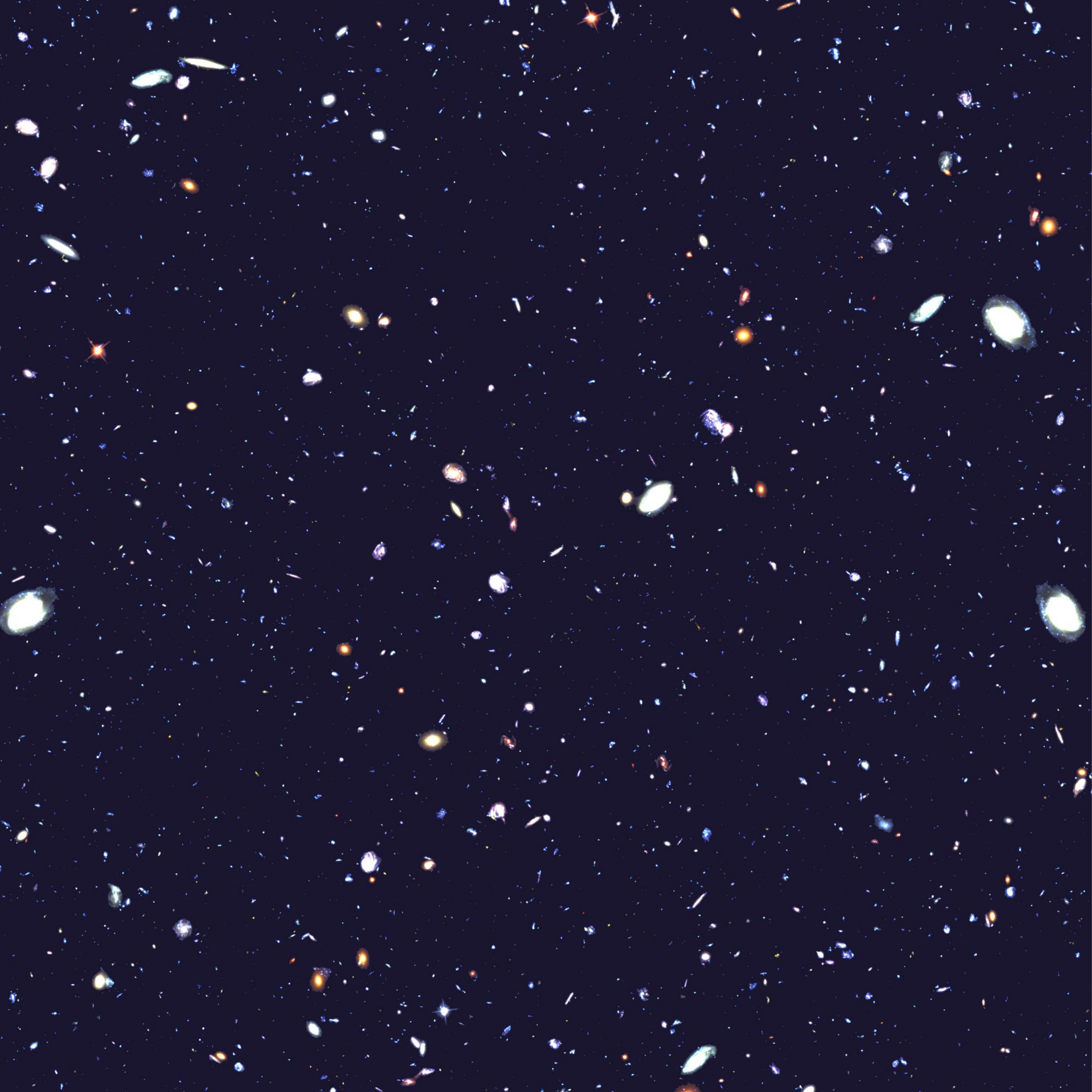}
	\caption{Example of simulated images of galaxies on the line of sight. 
	The size of the field of view and the resolution are the same as the Figure~\ref{fig:mem_imgs}. 
}
	\label{fig:los_imgs}
\end{figure}

To build an image of the unlensed light cone, we consider the galaxies
within the HUDF, as described in Section~\ref{sec:performing}.
Performing random sampling with replacement, we obtain a sample of
galaxies for a given field of view.  The galaxy number density of the
produced image is the same as HUDF, and the image is originally built
at the native sampling of the processed HUDF images of 0''.03 per pixel. Galaxies are pinned to the image at random positions and orientations---i.e., no correlation functions are built into the distribution. 
Finally, the complete image of the entire field of view is rebinned to
the requested pixelization (typically 0''.09 per pixel for images
discussed here), taking care to ensure conservation of total flux.
A sample image of simulated galaxies along the line of sight is shown in Figure~\ref{fig:los_imgs}.

\subsection{Stacking the Image Components}
\label{sec:stack}

Stacking the image of lensed galaxies, the image of galaxies on the
line of sight, and the image of member galaxies produces the final
realistic lensed image containing all the important extra-galactic
components. Care must be taken here to ensure the appropriate scaling
of each sub-image, in each filter. The HUDF data, for example, is not
in the identical filters as the RCS2 images from which the cluster
member photometry is derived, and the photometric zeropoints of each
subimage are not necessarily identical. Moreover, neither are likely
perfectly matched to the filter set of the real survey images for
which matching simulated images are being constructed. In the current
realisation of the pipeline, this aspect of the final image
construction is handled simplistically, since the image sub-sets used
are similar enough that simple bulk scalings, applied at the image
level, are sufficient to harmonize all the image subsets before
summing them. We proceed simply by adjusting all input image
zeropoints to that of the output simulated image in each desired
filter, adjusting for both overall sensitivity, and filter
shifts. 

Filter differences---for example with the HUDF F435W filter as input, being mapped into a desired g-band output image----are accounted for by applying a bulk scaling to the entire image that is based on the typical color ($g-F435W$) for galaxies in the HUDF field of view. In principle, when shifting the HUDF data
into other even nearby bandpasses, one should account for the spectrum
and redshift of each galaxy individually, as modest shifts in bandpass
may have quite large effects on flux if, for example, one is sampling
Lyman-break galaxies. However, it is worth noting that the typical lensed
galaxy is at z$\sim$2 \citep{Bayliss2012}, for which the above simple
(and fast) technique is likely sufficient. 

\begin{figure}
	\centering
	\includegraphics[width=0.475\textwidth]{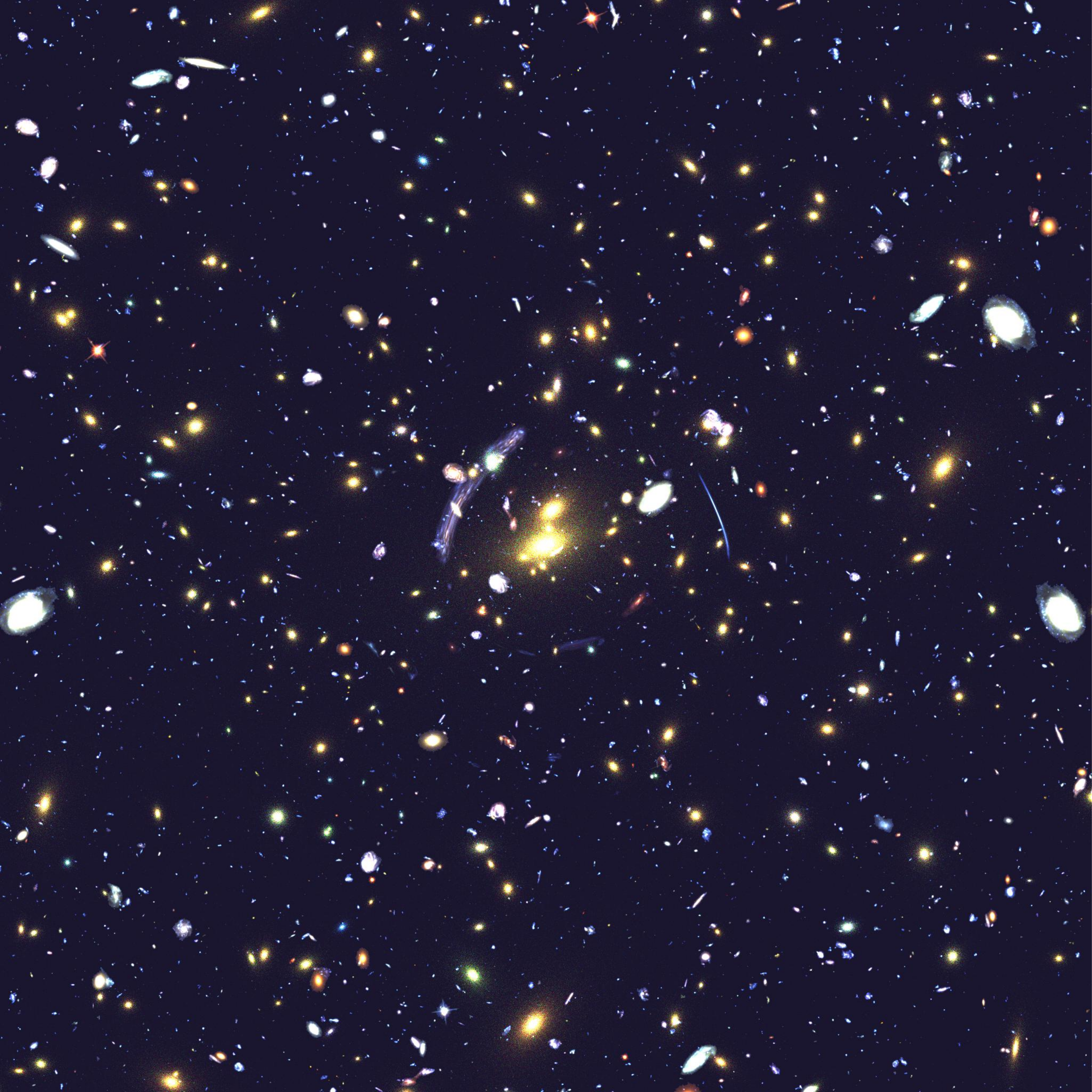}
	\caption{Example simulated image with all components
          added except bright stars.  The size of the field of view and the
          resolution are the same as the Figure~\ref{fig:mem_imgs}.
          This image is effectively a stack of
          Figures~\ref{fig:lensed_imgs}, \ref{fig:mem_imgs}, and
          \ref{fig:los_imgs}.  }
	\label{fig:all_imgs}
\end{figure}

Note also that the time-saving strategy employed in the
ray-tracing means that the central portion of the stacked image is actually
over-populated with distant (i.e., more distant than the lens)
galaxies. These galaxies are generally very faint---most too faint
to appear in any significant number in typical ground-based
imaging. There is no discernible over-density in a single image even
when not degraded to ground-based seeing and depth---see
Figure~\ref{fig:all_imgs}. However, one cannot measure the expected
de-magnification depletion behind massive lenses using these images;
for such a measure, or weak lensing measurements in general, a larger
field ray-trace must be adopted.

\subsection{Adding Bright Stars}
\label{sec:brightstar}
Bright stars are added to the simulated image in two steps. The
majority of stars---i.e., those which can be described as a point
source convolved with a model PSF---are drawn from an analysis of the
image set to be simulated. In practice, in a large set of survey data,
this information would be drawn from photometric catalogs that span
the many images of the survey. Note that proper star-galaxy
separation, which can be a significant issue in ground based data, is
not worrisome in this context; at the depths where this becomes
problematic, galaxy counts wholly dominate over star counts at
high galactic latitudes. From a visual perspective, ensuring the inclusion of stars
that are obviously point sources, and hence bright enough to be
robustly isolated in analysis of ground-based images, is sufficient.
For the example images detailed in Section~\ref{sec:results} below, the majority of
bright stars have been added as single pixel point sources to the
stacked image above, using data drawn from a catalog of stars
constructed by measuring unresolved sources in the real image these
examples are designed to replicate.

\begin{figure}
	\centering
	\includegraphics[width=0.475\textwidth]{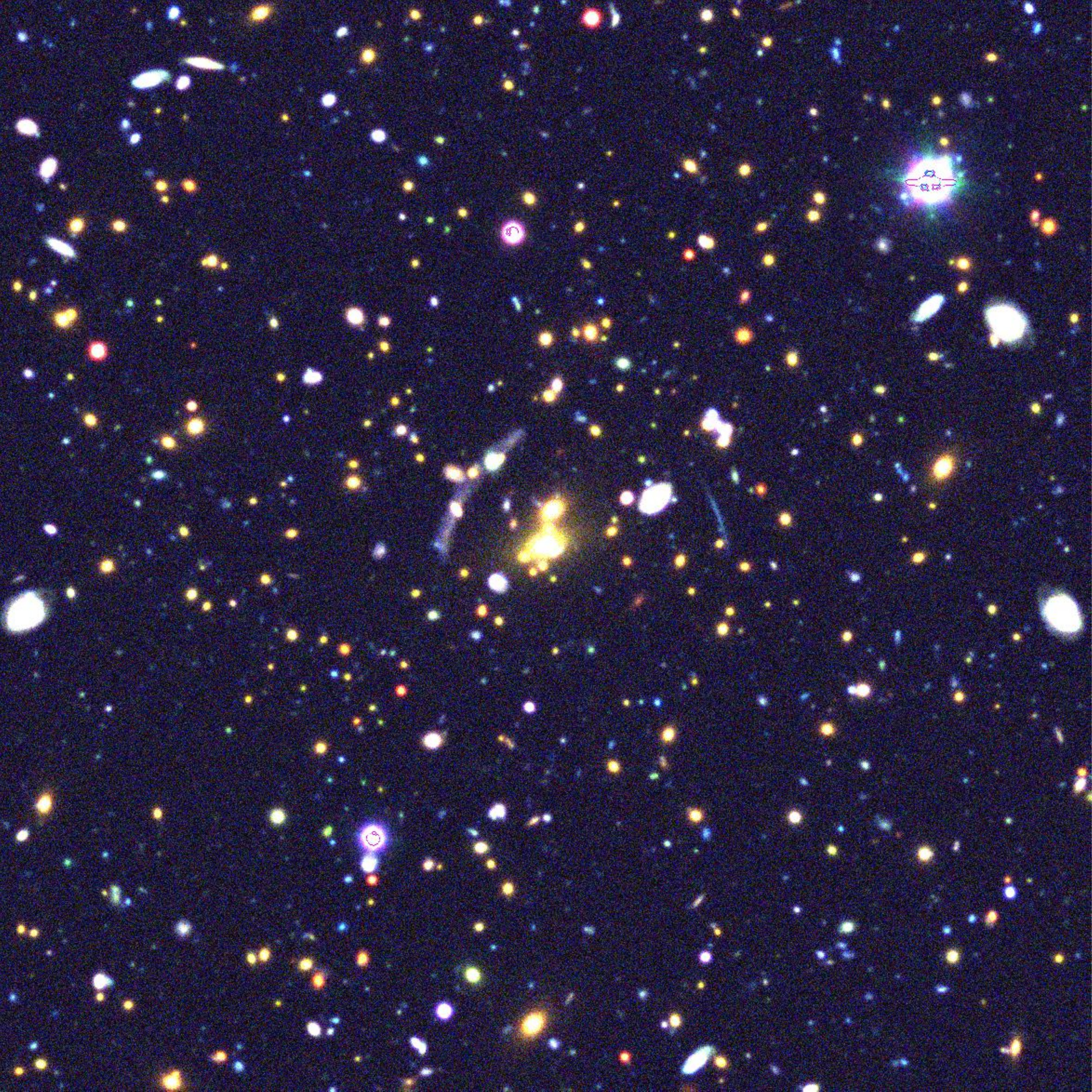}
	\caption{Example of a final simulated image after
          "observing"---i.e., convolution with an appropriate PSF,
          rebinning, and addition of noise and bright stars.  This
          image completes the pipeline sequence illustrated in Figures
          2-6.
}
	\label{fig:final_imgs}
\end{figure}

For the brightest stars---i.e., those for which the PSF becomes
visibly complex due to optical effects such as scattered light and
ghosting, and/or detector effects such as charge bleeding---the
pipeline uses a different strategy. Such stars are added {\it after} the
stacked image is convolved with a nominal telescope PSF (see
Section~\ref{sec:psf} below) by drawing on a catalog of image examples
taken from the real imaging data that the pipeline is
simulating. Image cutouts of bright stars are grafted in the simulated
images using a ragged boundary that is undetectable in the final image
product. This boundary is essentially defined algorithmically as a
noisy version of a low level isophote constructed by thresholding the
star image on a stack of all input images (typically three, if one is
simulating a suite of color images). Several examples of this can be
seen in Figure~\ref{fig:final_imgs}.

\subsection{Convolution with a PSF and Rebinning}
\label{sec:psf}

The entire image stack is next convolved with a PSF model. A simple
Moffat function \citep{Moffat1969} is used as the model. The normalized Moffat
profile is a two-parameter function of the form
\begin{equation}
  I(r)=\left[1+\left(\frac{r}{\alpha}\right)^2\right]^{-\beta},
 \label{eq:moffat}
\end{equation}
\noindent where $\alpha$ to first order describes the core of the PSF
and $\beta$ the wings, and the oft-used full-width-at-half-maximum
(FWHM) value of such a profile is given by 
\begin{equation}
\rm{FWHM} = 2\alpha\sqrt{2^{1/\beta} - 1}.
 \label{eq:fwhm}
\end{equation}

\noindent One of the necessary elements of simulating data
realistically is to allow for both the variation and correlation in
PSFs both within an image location between filters, and across a
dataset of multiple locations. We achieve this in the most
straightforward manner possible, by measuring the Moffat parameters
for each image in a dataset, and then drawing sets of PSF descriptions
(a set being all three filters, in the case of a typical color image)
from the same location in the survey data, but otherwise at
random. Once convolved with the appropriate PSF, each image is also
re-binned, conserving flux, to the final desired pixel scale, which is
set by the imaging data being simulated.

Care must be taken in the PSF convolution to use a large enough
convolution kernel such that the brightest image elements---generally,
point sources---do not show any significant edge effects from the
convolution box, once appropriate image noise is added (see
below). This is determined experimentally for any simulated image
set. It is desirable to keep the convolution box small for rapidity of
calculation; how one balances this against the typical PSF and noise
properties of a particular dataset to be simulated is best determined
at run time. 

\subsection{Adding Noise}
\label{sec:noise}
The final step in producing the simulated images is to add noise. In
PICS, the amount and character of noise added
to each image is guided by the noise statistics of the real data being
simulated. The baseline noise added is simply Poisson noise, with
amplitudes set by the real data. Like in the PSF convolution, this is
done by drawing a set of noise levels jointly across all filters under
consideration, to ensure the strong correlations that often exist
between filters in an imaging survey are sampled (in particular if
that imaging survey is taken in a mode where all filters are imaged
near-simultaneously at a given sky location). The default algorithm
actually selects both the PSF values and the sky noise levels from the
same location in the real imaging survey, as there may also be
correlations between sky levels and PSFs in real data.

Beyond simple Poisson noise, the pipeline also adds other noise
elements, which are crafted to match the specifics of the particular
data being simulated. Examples can be seen in the images in
Section~\ref{sec:results} below, and are discussed in detail there.

\section{Results}
\label{sec:results}
The SPT cluster sample \citep{Bleem2015} contains dozens of newly
discovered strong lenses. The optical imaging follow-up data used in
Bleem et al.\ is heterogeneous by optical imaging survey standards, as
it is a collection of pointed follow-up observations taken with a
variety of telescopes used in a wide range of observing
conditions. These real data thus serve as an excellent point of
comparison for further demonstrating the image simulation pipeline
described in detail above.

We have chosen for this purpose three example strong lensing SPT
clusters that are not representative of the typical SPT followup data
but rather illustrate the range of such data, and also span the
cluster redshift range where the bulk of cluster strong lensing
signals are expected for massive clusters \citep{Hennawi2008}. The
details of these three clusters, and the optical observations of them,
are given in Table~\ref{tab:spt}. In brief, the lowest redshift
system, SPT-CL~J2325-4111 (ACO S1121 \citet{Abell:1989mu}), 
is a particularly massive and optically-rich cluster, and the data used here are from the CTIO 4~m
telescope\footnote{http://www.ctio.noao.edu/noao/content/victor-blanco-4-m-telescope}
and MOSAIC-II camera.
These data are modestly deep but were taken under conditions of
extremely poor seeing. 

SPT-CL~J0307-5042, at a moderate redshift,
illustrates a strong-lensing system with several strongly-lensed
sources, and excellent data, i.e., long integrations with the MegaCam
imager \citep{mcleod06} at the 6.5 m Magellan II Clay
telescope\footnote{https://obs.carnegiescience.edu/Magellan}, taken
with seeing well less than one arcsecond. The highest redshift system,
SPT-CL~J0142-5032, is a cluster strong lens showing a single bright
giant arc, imaged quite shallowly using the 1~m Swope
telescope\footnote{https://obs.carnegiescience.edu/swope}. This
combination of shallow imaging, a small telescope, and a higher
redshift cluster renders all but the brightest cluster members
near-invisible, and is illustrative of a cluster detected at the sensitivity limit of a ground-based wide-field
imaging survey. 
Each of these real clusters also illustrates some
potential limitations of our methodology; these are discussed further
in Section~\ref{sec:discussion}. The simulated halos chosen to
approximate these real SPT strong lenses are each within 0.02 in
redshift of the real systems, and differ by less than 10\% from the
measured mass (i.e., well within the photometric redshift and mass
uncertainties). All three are taken from the Outer Rim simulation
described in Section~\ref{sec:cosmosim}.

\begin{table*}[t]
\centering
\resizebox{1.0\textwidth}{!}{
\begin{minipage}{1.0\textwidth}
\centering
\begin{tabular}{|l| cclcl |} 
\hline 
Cluster Name & Redshift & M$_{500} h_{70}^{-1}M_\odot$ & Telescope \& Camera (Pixel Scale)& Filters & Seeing(FWHM in '') \\ \hline \hline 
SPT-CL~J2325-4111 & 0.36 & 7.55$\times10^{14}$ & CTIO~4m+MOSAIC~II (0''.3)     & $g,r,i$ & 2.10,1.95,1.89  \\ 
SPT-CL~J0307-5042 & 0.55 & 5.26$\times10^{14}$ & Magellan~6.5m+MegaCam (0''.16) & $g,r,i$ & 0.90,0.47,0.73  \\ 
SPT-CL~J0142-5032 & 0.73 & 5.75$\times10^{14}$ & Swope~1m+SITe3 (0''.50) & $V,R,I$ & 1.30,1.16,1.14 \\ \hline \hline
\end{tabular}\par
\bigskip
\caption[SPT Strong-Lensing Comparison Clusters.]{Properties of the Clusters and Imaging Data, from \cite{Bleem2015}.}
\label{tab:spt} 
\end{minipage}}
\end{table*}

Figures~\ref{fig:swopefake} and \ref{fig:swopereal} illustrate
the shallowest images, from the Swope 1~m telescope. In each case we
leave it here as an exercise for the reader to determine which of the
two images is the real cluster, and which the simulation using the
pipeline tuned to match such data. An answer key is provided in
Section~\ref{sec:discussion}.  Figures~\ref{fig:mosaicreal} and
Figures~\ref{fig:mosaicfake} illustrate data from the CTIO 4~m
telescope. Finally, Figures~\ref{fig:megafake} and
Figures~\ref{fig:megareal} illustrate data from the 6.5~m Magellan
telescopes.  
 
Figures~\ref{fig:mosaicreal}-\ref{fig:megareal} demonstrate noise
window-paning due to observations with a mosaic camera, which results
in cross-hatched stripes of enhanced noise at inter-chip gaps, or even
small regions with no
data. Figures~\ref{fig:swopefake}-\ref{fig:mosaicfake} demonstrate two
forms of chip defects and cosmic rays. In
Figures~\ref{fig:swopefake}-\ref{fig:swopereal} these defects are
compact, but the real image has been processed using SWarp \citep{Bertin2002} to shift to a final output. This results in a
slight ringing in the image around sharp features. This effect is
captured in the image simulation by first creating a set of template
blank images with cosmic rays using IRAF's {\it mknoise} function, tuned
to match the statistics of the cosmic rays seen in the real data, and
then sampling and interpolating sub-images from that template set in
order to make instances of cosmic rays that include the subtler images
features induced by the processing of the real
data. Figures~\ref{fig:mosaicreal}-\ref{fig:mosaicfake} conversely
show how the pipeline adds more complex cosmic ray tracks. In this
case the real data being matched shows a small number of cosmic ray
tracks that traverse many pixels and appear `worm-like'. The same
basic algorithm that was used to add the brightest star images to the
simulated images is used in this case; actual images of these cosmic
rays are extracted from the real data, and added to the simulations.

\section{Discussion}
\label{sec:discussion}

\begin{figure*}
	\centering
	\includegraphics[width=1.0\textwidth]{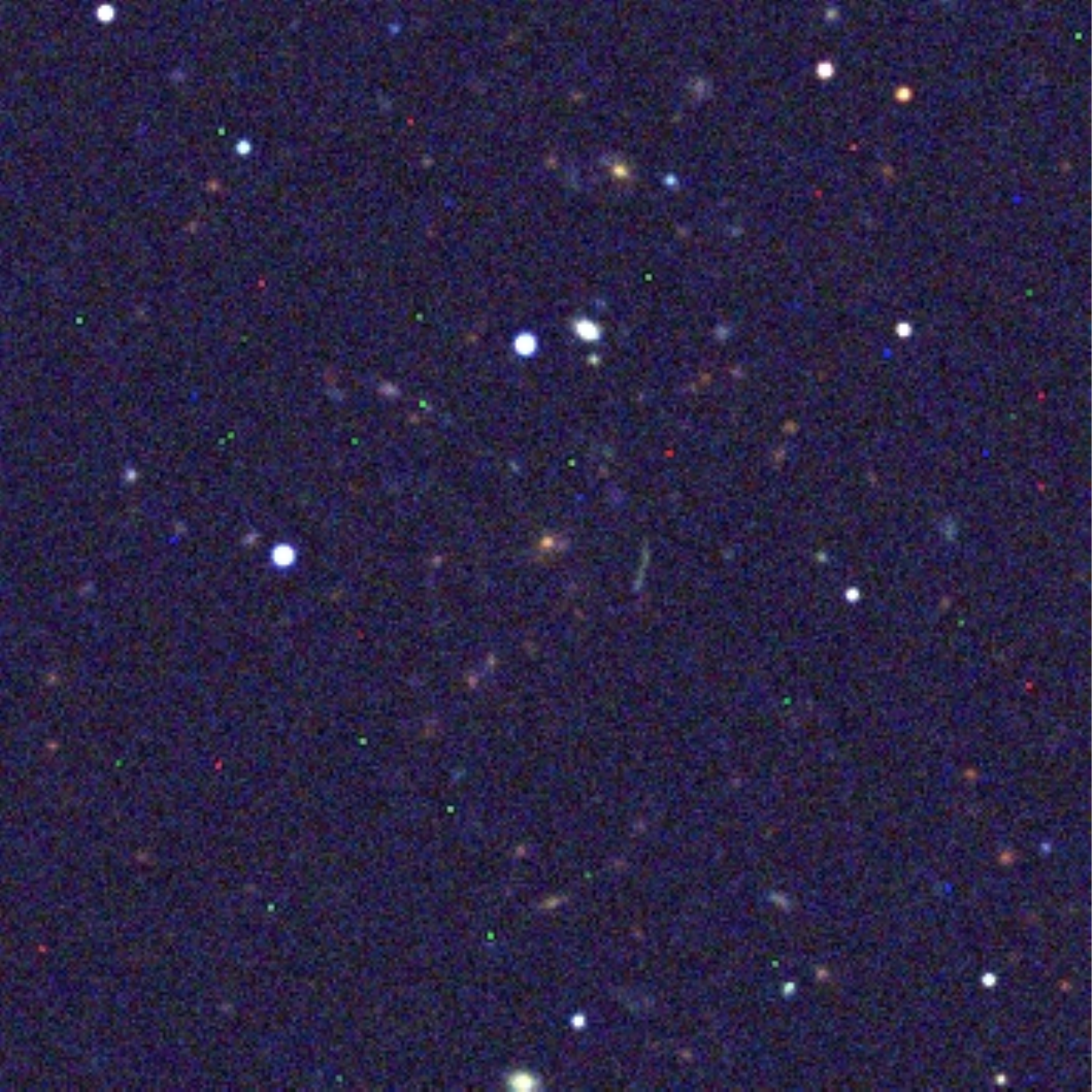}
	\caption{Example Swope 1~m observed strong lens from the SPT cluster sample, either a simulated or real image. An answer key is provided in Section~\ref{sec:discussion}.
 }
	\label{fig:swopefake}
\end{figure*}

\begin{figure*}
	\centering
	\includegraphics[width=1.0\textwidth]{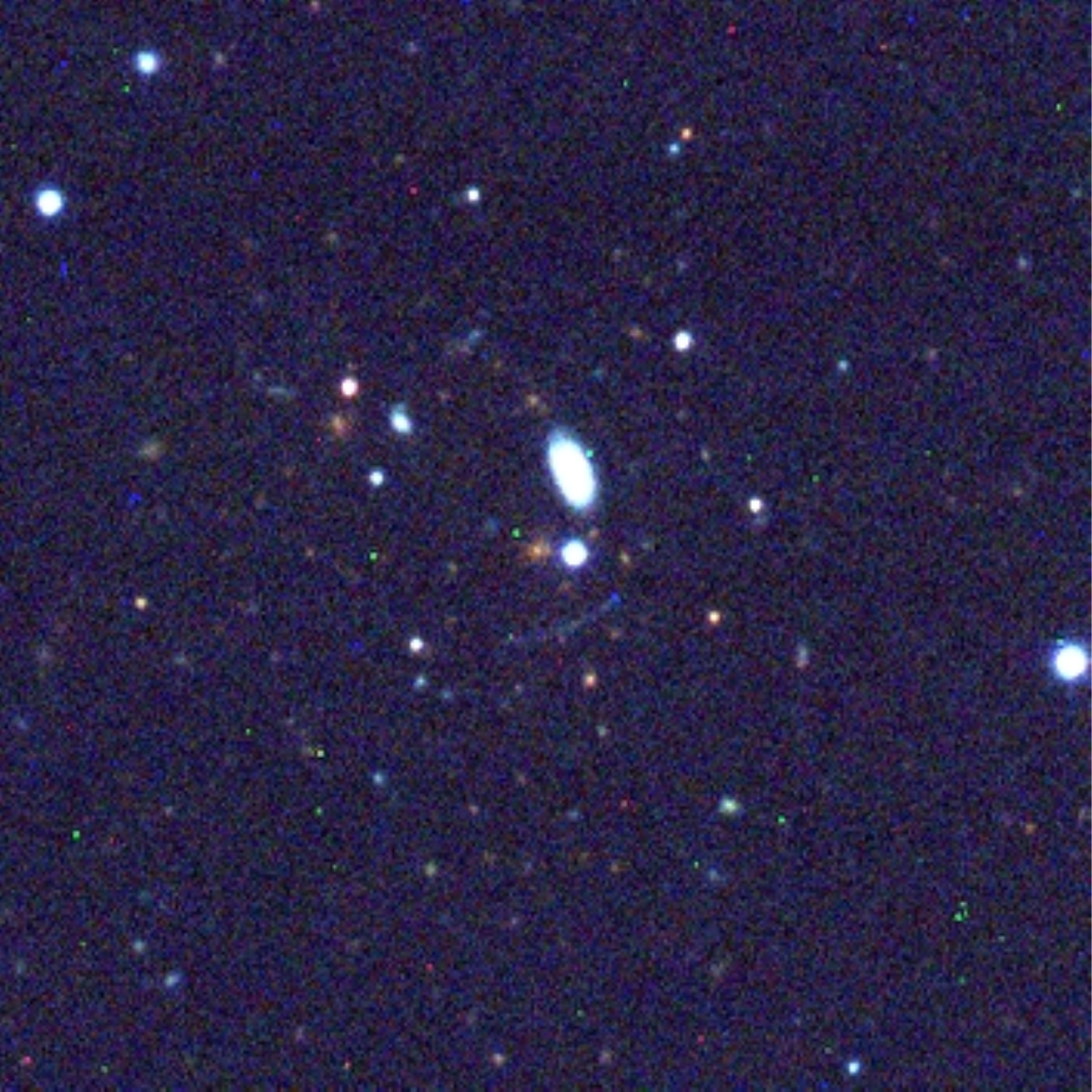}
	\caption{Example Swope 1~m observed strong lens from the SPT cluster sample, either a simulated or real image.
 }
	\label{fig:swopereal}
\end{figure*}

\begin{figure*}
	\centering
	\includegraphics[width=1.0\textwidth]{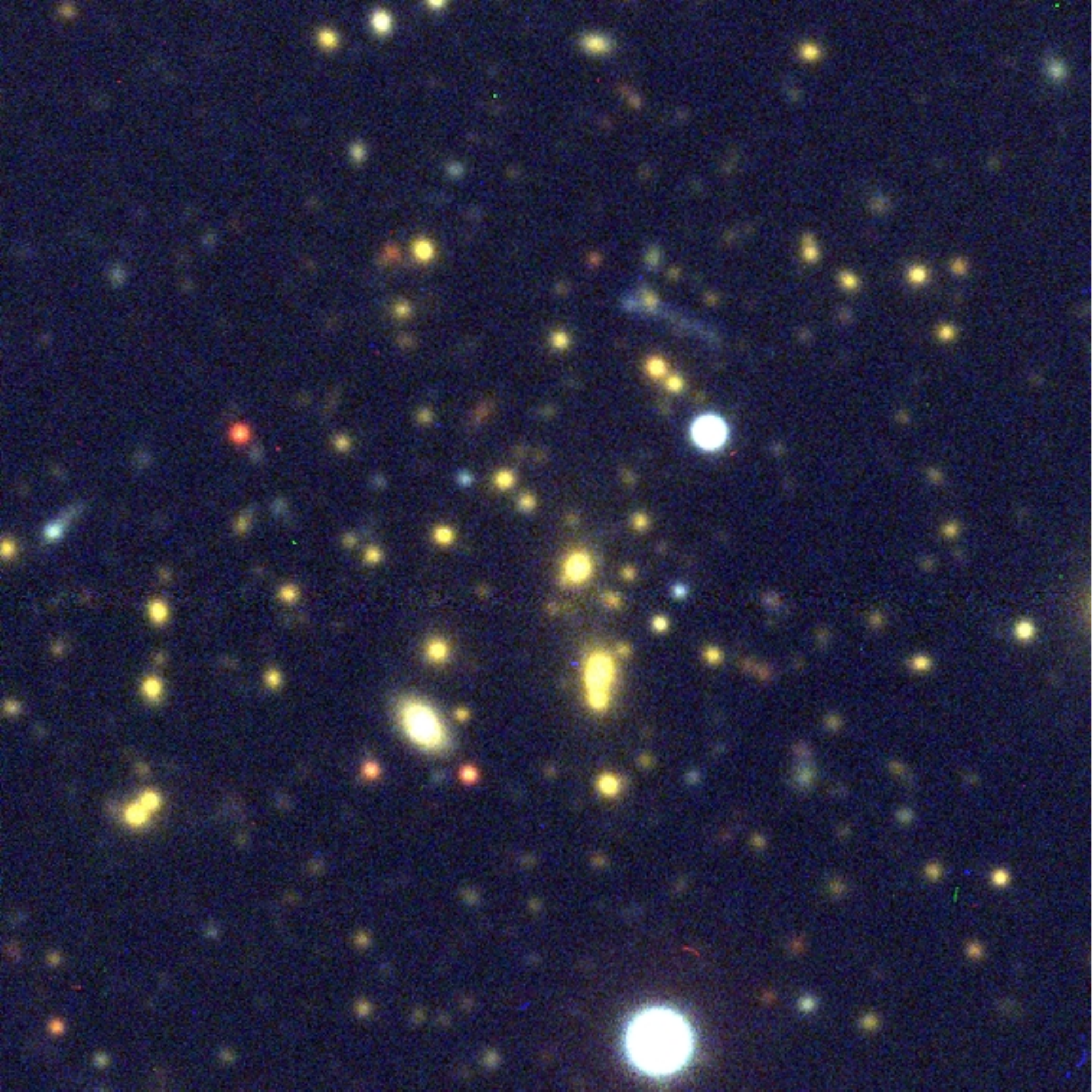}
	\caption{Example CTIO 4~m observed strong lens from the SPT cluster sample, either a simulated or real image.
 }
	\label{fig:mosaicreal}
\end{figure*}

\begin{figure*}
	\centering
	\includegraphics[width=1.0\textwidth]{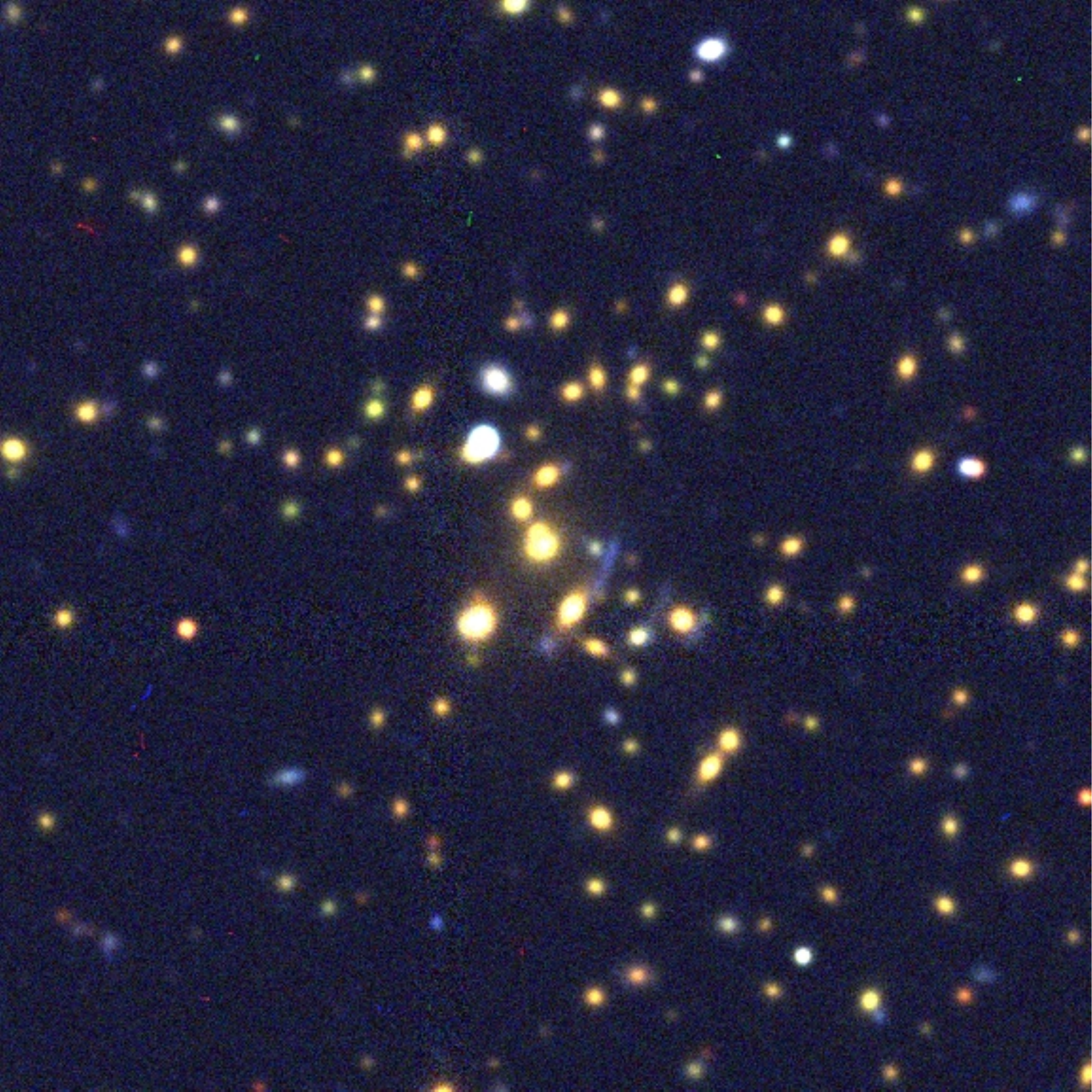}
	\caption{Example CTIO 4~m observed strong lens from the SPT cluster sample, either a simulated or real image.
 }
	\label{fig:mosaicfake}
\end{figure*}

\begin{figure*}
	\centering
	\includegraphics[width=1.0\textwidth]{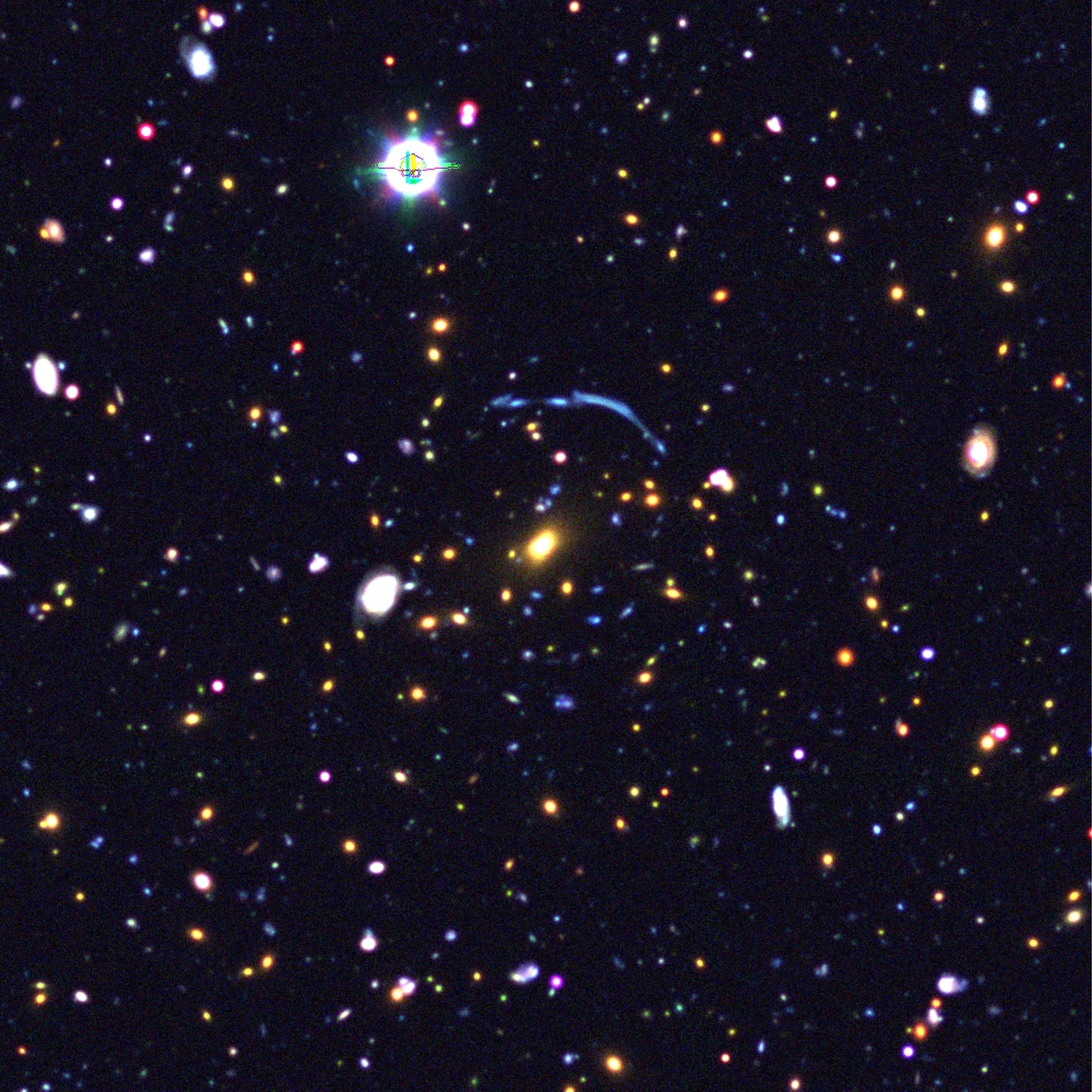}
	\caption{Example Magellan 6.5~m observed strong lens from the SPT cluster sample, either a simulated or real image.
 }
	\label{fig:megafake}
\end{figure*}

\begin{figure*}
	\centering
	\includegraphics[width=1.0\textwidth]{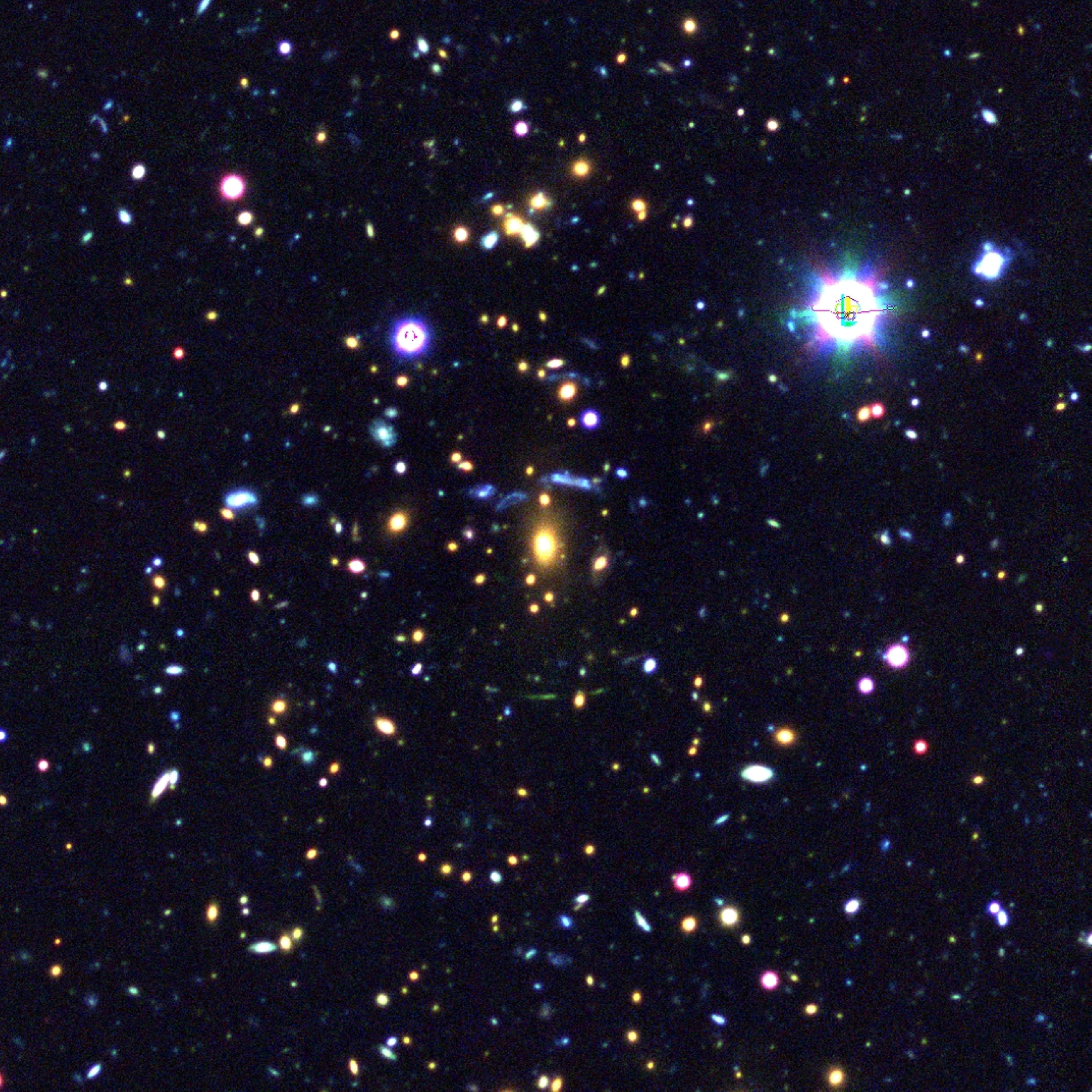}
	\caption{Example Magellan 6.5~m observed strong lens from the SPT cluster sample, either a simulated or real image.
 }
	\label{fig:megareal}
\end{figure*}

PICS produces simulated images of
strong lensing for halos over a wide range in
mass. We have focused our attention in this first paper on
demonstrating the efficacy of this pipeline for the most massive
strong lenses---i.e., massive clusters such as the recent sample from
the SPT \citep{Bleem2015}. Searches for strong lensing in imaging data---at 
least at optical wavelengths---have often proceeded by visual
searches
\citep[e.g.,][]{Gladders2003, Hennawi2008, Bayliss2011b, Wen2011, Stark2013, Marshall2015, More2016}
as the human eye offers remarkable sensitivity to complex patterned
signals in the low signal-to-noise regime. The aim of PICS is thus to produce rigorous strong-lensing calculations, by
ray-tracing extensive N-body simulations, and to couple these calculations
to a suite of algorithms capable of simulating the entire light cone
of an observation. These light cones can then be `observed' to produce realistic
simulated images which are essentially indistinguishable from real telescope data even to the expert eye. Figures~\ref{fig:swopefake}-\ref{fig:megareal}
give three paired examples of the outcomes:
Figures~\ref{fig:swopereal},\ref{fig:mosaicreal},\ref{fig:megareal}
are real SPT strong lenses, and
Figures~\ref{fig:swopefake},\ref{fig:mosaicfake},\ref{fig:megafake} the matched
simulated images. This last step - the production of simulated images that are
indistinguishable from real observations - will be important in the coming
decade as we move from samples of tens or hundreds of lenses to thousands of
systems. In such a regime, careful treatment of selection effects, achievable
likely only by passing simulated data through the same algorithms as applied to
real data, requires highly realistic image simulations.

A further goal is to enable the production of such images in large
numbers with a reasonable computational burden; exactly what
constitutes `reasonable' in this context is framed by the
computational cost of the input simulations, and the surveys and
imaging data to which such simulations are compared. Regardless, it is
instructive to consider the computational cost for a single image,
such as shown in Figures~\ref{fig:swopefake}-\ref{fig:megareal}. 
PICS images can be produced on workstations or on parallel supercomputers. Timing examples of these two use cases are: 
\begin{itemize}

\item a test based on producing a single image, from a pipeline
  instance executing on a single thread on one core of a data analysis workstation, a.k.a. `DataStar'. This
  computer, used for local and realtime data analysis tasks, provides
  24 dual-threaded Xeon E5-2620 cores running at 2.0 GHz, with ample
  memory resources.

\item a test based on producing 1000 images, one from each of 1000
  pipeline instances, running in parallel on a total of 1000 cores on
  the NERSC supercomputer
  `Edison'. Edison
  is a Cray XC30 supercomputer with 133,824 cores, where each core
  is an Intel ``Ivy Bridge'' 2.4Ghz processor, again with ample memory
  resources. Edison represents a highly parallelized compute
  environment, suitable for considering resources for large production
  runs of this image simulation pipeline.
\end{itemize}

The lens density estimation takes a little less than half the time, roughly equal time is spent on ray-tracing, generating member galaxy images, and galaxy images along the line of sight. The computational cost of the deflection angle field and `observing' the image stack is sub-dominant.

Note that the calculations directly related to the strong lensing take
about 60\% of the total time, and that at this pace one could produce
about 20,000 simulated images on a workstation scale computer like
DataStar in a week of runtime. The individual calculations on
the Edison are slower, however, a one week
run using only a modest 4\% of the available power of that machine
would produce $\sim$1 million simulated strong lensing images, such as
one might carry out as an input to the analysis of a survey covering a large
fraction of the entire sky.

The primary purpose of these simulated images is to enable the study of arc
statistics, by providing a robust link between real data and
simulation predictions. A second paper in this series will explore
this application in detail for the SPT cluster sample. However, it is
worth noting that the pipeline itself is sufficiently flexible to
enable a number of other studies. Elements of this pipeline have been
used to study lensing effects on the morphologies of background source
galaxies \citep{florian2015a} and the utility of the Gini
coefficient as a constraint when identifying image families in strong
lesing systems \citep{florian2015b}. In the longer term,
development of automated strong-lens-finding algorithms
\citep[e.g.,][]{Seidel2007, Marshall2009, Joseph2014, Gavazzi2014, Xu2016} will be enhanced by
the ready availability of large libraries of realistic test images
against which algorithmic advances may be rigorously tested. Moreover, only
minor tweaks to the pipeline as laid out here, or changes to the
background source, will enable full simulation of weak
lensing in galaxy clusters, including lensing of the cosmic microwave background by
clusters. Changes to the input mass field, in order to better
accommodate the effects of bayornic structures in individual galaxies,
will also enable studies of galaxy-galaxy lensing.

PICS has, however, some limitations
which must be considered in the context of some applications or real
datasets. Several issues arise due to the use of the HUDF as the input
data for ray-tracing source planes. The first issue is the angular
area---and cosmological volume---covered by these data. For the
brightest galaxies at any redshift, the HUDF does not provide
well-sampled statistics with negligible cosmic variance. Thus, it fails
to include the occasional bright low-z galaxy which does appear in
real data along lines of sight to known strong lenses: see, for
example, the bright, resolved late-type galaxy adjacent to the strong
lens in Figure~\ref{fig:swopereal}. This same limitation also makes
the statistics of only the very brightest arcs---those that come from
systems with highly magnified and intrinsically bright
galaxies\cite[e.g.,][]{Wuyts2012}---questionable when using only the HUDF as a
source
input.  Moreover, if one is simulating a field larger than the HUDF,
repetition of individual galaxies in the unlensed light-cone outputs
is unavoidable, and can produce visual hints that the image in
question is simulated: see, for example, the three instances of the
same recognizable spiral galaxy that appear on the left and right
edges of the light cone example images in
Figure~\ref{fig:los_imgs}. All three of these related issues can (and will) be
addressed in a future iteration of this
pipeline by both grafting wider-field multi-band HST observations (which,
perforce, will be shallower) together with the HUDF and by sampling the
source planes with a `wedding-cake' strategy that minimizes the
effects of the HUDF's small field of view while still exploiting its
depth, image quality, and extensive characterization in the literature.

However, even the HUDF, or the yet deeper refined dataset the HST eXtreme
Deep Field \citep{Illingworth2013}, fail fundamentally to address one limitation
of using real images as inputs to strong-lensing ray-tracing: namely 
that the strongly-lensed images provide access to smaller spatial
scales in the source plane due to magnification. This limitation is
insignificant when simulating most ground-based data, but at spatial
resolutions of $\sim$0.5'' or better, such as when comparing the best
ground-based imaging to equivalent simulations (see
Figure~\ref{fig:megareal} compared to Figure~\ref{fig:megafake}), the
lack of finer spatial detail in the HST-based input data starts to
become apparent. This effect will be obvious when simulating
high-resolution space-based observations of strong lensing. To address
this requires the generation of spatial information in the source plane at scales finer
that the best available data. In some applications, redshifting images
of low redshift galaxies to high redshift provides the necessary
inputs \citep[e.g.,][]{florian2015a}. However, for large-scale survey
predictions, a more robust method that adds spatial complexity to
deep extant HST images is likely needed. One approach may be to use
rapid galaxy-image-simulation tools such as GAMER
\citep{Groeneboom2014} to complexify HST images.

Multiple lens-planes are also enabled in the image simulation
pipeline. For large sample arcs statistics, some claim that the
effects of structure on the line of sight is modest \citep[$\lesssim
  7\%$,][] {Hennawi2007}, though \cite{Puchwein2009} assert a boost of
$10\%-25\%$ if the effects of additional structures along the
line of sight are included, and \cite{Wambsganss2005} note that the
magnitude of the boost is a strong function of source redshift.
Doubtless, for high accuracy lensing modeling
\citep{Inoue2012, DAloisio2014} and time delay measurements of a single
lensing system, the lensing effects of the structure on the line of
sight should be taken into account \citep{McCully2014, Bayliss2014},
and simulations of gravitational lensing with multiple planes are
needed.  We will investigate the influence of structures on the line
of sight on gravitational lensing systematically in a future paper,
using the multiple lens-plane capabilities of the code presented here.

In this work, a gravity-only N-body simulation is used, in part
because the primary aim of this paper is to describe the image
simulation pipeline, rather than to discuss in detail issues of resolution,
N-body versus hydrodynamical simulations \citep{Killedar2012b}, galaxy modeling in clusters, etc., when computing strong lensing
statistics, issues to which we will return in future work. Observationally, \cite{Newman2013} show that the total density profile of the central
galaxy and the dark matter halo in massive clusters can be described
by an NFW profile \citep{1996ApJ...462..563N}, implying that the presence of baryonic matter does
not significantly change the original form of the density profile in
massive cluster-scale halos. Regarding substructure, as shown by
\cite{nagai2005} using hydrodynamical simulations, baryons do affect the
subhalos at scales smaller than the virial radius ($\lesssim 0.2r_{vir}$) but not
significantly at larger radii. Lensing effects due to
baryonic matter in the member galaxies cannot ultimately be ignored,
because gravitationally lensed arcs which are produced by member
galaxies are observed
\citep{Halkola2006,Sand2008,Newman2009}. Regardless, the presence or
lack of disagreement between simulations and observations of strong
lensing remains an open question and we anticipate that the simulation
toolkit described here will add usefully to that ongoing
discussion. 

\acknowledgements

This research was funded in part by the Strategic Collaborative
Initiative administered by the University of Chicago's Office of the
Vice President for Research and for National Laboratories. Argonne
National Laboratory's work was supported under the U.S. Department of
Energy contract DE-AC02-06CH11357. This research used resources of the
ALCF, which is supported by DOE/SC under contract DE-AC02-06CH11357
and resources of the OLCF, which is supported by DOE/SC under contract
DE-AC05-00OR22725. Some of the results presented here result from
awards of computer time provided by the ASCR Leadership Computing
Challenge (ALCC) programs at Argonne and Oak Ridge (ALCF and
OLCF). This work was also supported in part by the Kavli Institute for
Cosmological Physics at the University of Chicago through grant NSF
PHY-1125897 and an endowment from the Kavli Foundation and its founder
Fred Kavli. We also thank the South Pole Telescope collaboration for
allowing the usage of the SPT cluster follow-up imaging shown in this
work. Finally, we enthusiastically thank the referee, Massimo
Meneghetti, for a particularly thoughtful and constructive review of
this paper, which served to signficantly enhance the quality of the
final manuscript.

\bibliographystyle{apj}
\bibliography{./ms.bib}

\begin{thebibliography}{}
\expandafter\ifx\csname natexlab\endcsname\relax\def\natexlab#1{#1}\fi

\bibitem[{{Abell} {et~al.}(1989){Abell}, {Corwin}, \& {Olowin}}]{Abell:1989mu}
{Abell}, G.~O., {Corwin}, Jr., H.~G., \& {Olowin}, R.~P. 1989, \apjs, 70, 1

\bibitem[{{Abraham} {et~al.}(2003){Abraham}, {van den Bergh}, \&
  {Nair}}]{Abraham2003}
{Abraham}, R.~G., {van den Bergh}, S., \& {Nair}, P. 2003, \apj, 588, 218

\bibitem[{{Barnes} \& {Hut}(1986)}]{barneshut}
{Barnes}, J., \& {Hut}, P. 1986, \nat, 324, 446

\bibitem[{{Bartelmann} {et~al.}(1998){Bartelmann}, {Huss}, {Colberg},
  {Jenkins}, \& {Pearce}}]{Bartelmann1998}
{Bartelmann}, M., {Huss}, A., {Colberg}, J.~M., {Jenkins}, A., \& {Pearce},
  F.~R. 1998, \aap, 330, 1

\bibitem[{{Bayliss}(2012)}]{Bayliss2012}
{Bayliss}, M.~B. 2012, \apj, 744, 156

\bibitem[{{Bayliss} {et~al.}(2011{\natexlab{a}}){Bayliss}, {Gladders}, {Oguri},
  {Hennawi}, {Sharon}, {Koester}, \& {Dahle}}]{Bayliss2011}
{Bayliss}, M.~B., {Gladders}, M.~D., {Oguri}, M., {et~al.} 2011{\natexlab{a}},
  \apjl, 727, L26

\bibitem[{{Bayliss} {et~al.}(2011{\natexlab{b}}){Bayliss}, {Hennawi},
  {Gladders}, {Koester}, {Sharon}, {Dahle}, \& {Oguri}}]{Bayliss2011b}
{Bayliss}, M.~B., {Hennawi}, J.~F., {Gladders}, M.~D., {et~al.}
  2011{\natexlab{b}}, \apjs, 193, 8

\bibitem[{{Bayliss} {et~al.}(2014){Bayliss}, {Johnson}, {Gladders}, {Sharon},
  \& {Oguri}}]{Bayliss2014}
{Bayliss}, M.~B., {Johnson}, T., {Gladders}, M.~D., {Sharon}, K., \& {Oguri},
  M. 2014, \apj, 783, 41

\bibitem[{{Beckwith} {et~al.}(2006){Beckwith}, {Stiavelli}, {Koekemoer},
  {Caldwell}, {Ferguson}, {Hook}, {Lucas}, {Bergeron}, {Corbin}, {Jogee},
  {Panagia}, {Robberto}, {Royle}, {Somerville}, \& {Sosey}}]{Beckwith2006}
{Beckwith}, S.~V.~W., {Stiavelli}, M., {Koekemoer}, A.~M., {et~al.} 2006, \aj,
  132, 1729

\bibitem[{{Behroozi} {et~al.}(2013){Behroozi}, {Wechsler}, \& {Wu}}]{rockstar}
{Behroozi}, P.~S., {Wechsler}, R.~H., \& {Wu}, H.-Y. 2013, \apj, 762, 109

\bibitem[{{Bertin} {et~al.}(2002){Bertin}, {Mellier}, {Radovich}, {Missonnier},
  {Didelon}, \& {Morin}}]{Bertin2002}
{Bertin}, E., {Mellier}, Y., {Radovich}, M., {et~al.} 2002, in Astronomical
  Society of the Pacific Conference Series, Vol. 281, Astronomical Data
  Analysis Software and Systems XI, ed. D.~A. {Bohlender}, D.~{Durand}, \&
  T.~H. {Handley}, 228

\bibitem[{{Bhattacharya} {et~al.}(2013){Bhattacharya}, {Habib}, {Heitmann}, \&
  {Vikhlinin}}]{Bhattacharya2013}
{Bhattacharya}, S., {Habib}, S., {Heitmann}, K., \& {Vikhlinin}, A. 2013, \apj,
  766, 32

\bibitem[{{Bleem} {et~al.}(2015){Bleem}, {Stalder}, {de Haan}, {Aird}, {Allen},
  {Applegate}, {Ashby}, {Bautz}, {Bayliss}, {Benson}, {Bocquet}, {Brodwin},
  {Carlstrom}, {Chang}, {Chiu}, {Cho}, {Clocchiatti}, {Crawford}, {Crites},
  {Desai}, {Dietrich}, {Dobbs}, {Foley}, {Forman}, {George}, {Gladders},
  {Gonzalez}, {Halverson}, {Hennig}, {Hoekstra}, {Holder}, {Holzapfel},
  {Hrubes}, {Jones}, {Keisler}, {Knox}, {Lee}, {Leitch}, {Liu}, {Lueker},
  {Luong-Van}, {Mantz}, {Marrone}, {McDonald}, {McMahon}, {Meyer}, {Mocanu},
  {Mohr}, {Murray}, {Padin}, {Pryke}, {Reichardt}, {Rest}, {Ruel}, {Ruhl},
  {Saliwanchik}, {Saro}, {Sayre}, {Schaffer}, {Schrabback}, {Shirokoff},
  {Song}, {Spieler}, {Stanford}, {Staniszewski}, {Stark}, {Story}, {Stubbs},
  {Vanderlinde}, {Vieira}, {Vikhlinin}, {Williamson}, {Zahn}, \&
  {Zenteno}}]{Bleem2015}
{Bleem}, L.~E., {Stalder}, B., {de Haan}, T., {et~al.} 2015, \apjs, 216, 27

\bibitem[{{Bolton} {et~al.}(2006){Bolton}, {Burles}, {Koopmans}, \&
  {Moustakas}}]{Bolton2006}
{Bolton}, A.~S., {Burles}, S., {Koopmans}, L.~V.~E.~and{Treu}, T., \&
  {Moustakas}, L.~A. 2006, \apj, 638, 703

\bibitem[{{Brada{\v c}} {et~al.}(2004{\natexlab{a}}){Brada{\v c}}, {Lombardi},
  \& {Schneider}}]{Bradac2004}
{Brada{\v c}}, M., {Lombardi}, M., \& {Schneider}, P. 2004{\natexlab{a}}, \aap,
  424, 13

\bibitem[{{Brada{\v c}} {et~al.}(2004{\natexlab{b}}){Brada{\v c}}, {Schneider},
  {Lombardi}, {Steinmetz}, {Koopmans}, \& {Navarro}}]{Bradac2004b}
{Brada{\v c}}, M., {Schneider}, P., {Lombardi}, M., {et~al.}
  2004{\natexlab{b}}, \aap, 423, 797

\bibitem[{{Coe} {et~al.}(2006){Coe}, {Ben{\'{\i}}tez}, {S{\'a}nchez}, {Jee},
  {Bouwens}, \& {Ford}}]{Coe2006}
{Coe}, D., {Ben{\'{\i}}tez}, N., {S{\'a}nchez}, S.~F., {et~al.} 2006, \aj, 132,
  926

\bibitem[{{Coe} {et~al.}(2013){Coe}, {Zitrin}, {Carrasco}, {Shu}, {Zheng},
  {Postman}, {Bradley}, {Koekemoer}, {Bouwens}, {Broadhurst}, {Monna}, {Host},
  {Moustakas}, {Ford}, {Moustakas}, {van der Wel}, {Donahue}, {Rodney},
  {Ben{\'{\i}}tez}, {Jouvel}, {Seitz}, {Kelson}, \& {Rosati}}]{Coe2013}
{Coe}, D., {Zitrin}, A., {Carrasco}, M., {et~al.} 2013, \apj, 762, 32

\bibitem[{{Cohn} \& {White}(2008)}]{cohnwhite}
{Cohn}, J.~D., \& {White}, M. 2008, \mnras, 385, 2025

\bibitem[{{Collett}(2015)}]{Collett2015}
{Collett}, T.~E. 2015, \apj, 811, 20

\bibitem[{{Dalal} {et~al.}(2004){Dalal}, {Holder}, \& {Hennawi}}]{Dalal2004}
{Dalal}, N., {Holder}, G., \& {Hennawi}, J.~F. 2004, \apj, 609, 50

\bibitem[{{D'Aloisio} {et~al.}(2014){D'Aloisio}, {Natarajan}, \&
  {Shapiro}}]{DAloisio2014}
{D'Aloisio}, A., {Natarajan}, P., \& {Shapiro}, P.~R. 2014, \mnras, 445, 3581

\bibitem[{{Dressler} {et~al.}(1997){Dressler}, {Oemler}, {Couch}, {Smail},
  {Ellis}, {Barger}, {Butcher}, {Poggianti}, \& {Sharples}}]{Dressler1997}
{Dressler}, A., {Oemler}, Jr., A., {Couch}, W.~J., {et~al.} 1997, \apj, 490,
  577

\bibitem[{{Dye} {et~al.}(2014){Dye}, {Negrello}, {Hopwood}, {Nightingale},
  {Bussmann}, {Amber}, {Bourne}, {Cooray}, {Dariush}, {Dunne}, {Eales},
  {Gonzalez-Nuevo}, {Ibar}, {Ivison}, {Maddox}, {Valiante}, \&
  {Smith}}]{Dye2014}
{Dye}, S., {Negrello}, M., {Hopwood}, R., {et~al.} 2014, \mnras, 440, 2013

\bibitem[{{Fasano} {et~al.}(2012){Fasano}, {Vanzella}, {Dressler}, {Poggianti},
  {Moles}, {Bettoni}, {Valentinuzzi}, {Moretti}, {D'Onofrio}, {Varela},
  {Couch}, {Kj{\ae}rgaard}, {Fritz}, {Omizzolo}, \& {Cava}}]{Fasano2012}
{Fasano}, G., {Vanzella}, E., {Dressler}, A., {et~al.} 2012, \mnras, 420, 926

\bibitem[{{Faure} {et~al.}(2009){Faure}, {Kneib}, {Hilbert}, {Massey},
  {Covone}, {Finoguenov}, {Leauthaud}, {Taylor}, {Pires}, {Scoville}, \&
  {Koekemoer}}]{Faure2009}
{Faure}, C., {Kneib}, J.-P., {Hilbert}, S., {et~al.} 2009, \apj, 695, 1233

\bibitem[{{Fedeli} {et~al.}(2007){Fedeli}, {Bartelmann}, {Meneghetti}, \&
  {Moscardini}}]{Fedeli2007}
{Fedeli}, C., {Bartelmann}, M., {Meneghetti}, M., \& {Moscardini}, L. 2007,
  \aap, 473, 715

\bibitem[{{Fedeli} {et~al.}(2008){Fedeli}, {Bartelmann}, {Meneghetti}, \&
  {Moscardini}}]{Fedeli2008}
---. 2008, \aap, 486, 35

\bibitem[{{Fedeli} {et~al.}(2010){Fedeli}, {Meneghetti}, {Gottl{\"o}ber}, \&
  {Yepes}}]{Fedeli2010}
{Fedeli}, C., {Meneghetti}, M., {Gottl{\"o}ber}, S., \& {Yepes}, G. 2010, \aap,
  519, A91

\bibitem[{{Florian} {et~al.}(2015){Florian}, {Gladders}, {Li}, \&
  {Sharon}}]{florian2015b}
{Florian}, M.~K., {Gladders}, M.~D., {Li}, N., \& {Sharon}, K. 2015, ArXiv
  e-prints, arXiv:1511.03594

\bibitem[{{Florian} {et~al.}(2016){Florian}, {Gladders}, {Li}, \&
  {Sharon}}]{florian2015a}
---. 2016, \apjl, 816, L23

\bibitem[{{Gao} {et~al.}(2009){Gao}, {Jing}, {Mao}, {Li}, \& {Kong}}]{Gao2009}
{Gao}, G.~J., {Jing}, Y.~P., {Mao}, S., {Li}, G.~L., \& {Kong}, X. 2009, \apj,
  707, 472

\bibitem[{{Gavazzi} {et~al.}(2014){Gavazzi}, {Marshall}, {Treu}, \&
  {Sonnenfeld}}]{Gavazzi2014}
{Gavazzi}, R., {Marshall}, P.~J., {Treu}, T., \& {Sonnenfeld}, A. 2014, \apj,
  785, 144

\bibitem[{{Gilbank} {et~al.}(2011){Gilbank}, {Gladders}, {Yee}, \&
  {Hsieh}}]{Gilbank2011}
{Gilbank}, D.~G., {Gladders}, M.~D., {Yee}, H.~K.~C., \& {Hsieh}, B.~C. 2011,
  \aj, 141, 94

\bibitem[{{Giocoli} {et~al.}(2012){Giocoli}, {Meneghetti}, {Bartelmann},
  {Moscardini}, \& {Boldrin}}]{Giocoli2012}
{Giocoli}, C., {Meneghetti}, M., {Bartelmann}, M., {Moscardini}, L., \&
  {Boldrin}, M. 2012, \mnras, 421, 3343

\bibitem[{{Gladders} {et~al.}(2003){Gladders}, {Hoekstra}, {Yee}, {Hall}, \&
  {Barrientos}}]{Gladders2003}
{Gladders}, M.~D., {Hoekstra}, H., {Yee}, H.~K.~C., {Hall}, P.~B., \&
  {Barrientos}, L.~F. 2003, \apj, 593, 48

\bibitem[{{Gould} \& {Yee}(2014)}]{Gould2014}
{Gould}, A., \& {Yee}, J.~C. 2014, \apj, 784, 64

\bibitem[{{Grillo} {et~al.}(2015){Grillo}, {Suyu}, {Rosati}, {Mercurio},
  {Balestra}, {Munari}, {Nonino}, {Caminha}, {Lombardi}, {De Lucia}, {Borgani},
  {Gobat}, {Biviano}, {Girardi}, {Umetsu}, {Coe}, {Koekemoer}, {Postman},
  {Zitrin}, {Halkola}, {Broadhurst}, {Sartoris}, {Presotto}, {Annunziatella},
  {Maier}, {Fritz}, {Vanzella}, \& {Frye}}]{Grillo2015}
{Grillo}, C., {Suyu}, S.~H., {Rosati}, P., {et~al.} 2015, \apj, 800, 38

\bibitem[{{Groeneboom} \& {Dahle}(2014)}]{Groeneboom2014}
{Groeneboom}, N.~E., \& {Dahle}, H. 2014, \apj, 783, 138

\bibitem[{{Habib} {et~al.}(2014){Habib}, {Pope}, {Finkel}, {Frontiere},
  {Heitmann}, {Daniel}, {Fasel}, {Morozov}, {Zagaris}, {Peterka}, {Vishwanath},
  {Lukic}, {Sehrish}, \& {Liao}}]{habib14}
{Habib}, S., {Pope}, A., {Finkel}, H., {et~al.} 2014, ArXiv e-prints,
  arXiv:1410.2805

\bibitem[{{Halkola} {et~al.}(2006){Halkola}, {Seitz}, \&
  {Pannella}}]{Halkola2006}
{Halkola}, A., {Seitz}, S., \& {Pannella}, M. 2006, \mnras, 372, 1425

\bibitem[{{Han} {et~al.}(2013){Han}, {Jung}, {Udalski}, {Sumi}, {Gaudi},
  {Gould}, {Bennett}, {Tsapras}, {Szyma{\'n}ski}, {Kubiak}, {Pietrzy{\'n}ski},
  {Soszy{\'n}ski}, {Skowron}, {Koz{\l}owski}, {Poleski}, {Ulaczyk},
  {Wyrzykowski}, {Pietrukowicz}, {OGLE Collaboration}, {Abe}, {Bond},
  {Botzler}, {Chote}, {Freeman}, {Fukui}, {Furusawa}, {Harris}, {Itow}, {Ling},
  {Masuda}, {Matsubara}, {Muraki}, {Ohnishi}, {Rattenbury}, {Saito},
  {Sullivan}, {Sweatman}, {Suzuki}, {Tristram}, {Wada}, {Yock}, {MOA
  Collaboration}, {Batista}, {Christie}, {Choi}, {DePoy}, {Dong}, {Hwang},
  {Kavka}, {Lee}, {Monard}, {Natusch}, {Ngan}, {Park}, {Pogge}, {Porritt},
  {Shin}, {Tan}, {Yee}, {{$\mu$}FUN Collaboration}, {Alsubai}, {Bozza},
  {Bramich}, {Browne}, {Dominik}, {Horne}, {Hundertmark}, {Ipatov}, {Kains},
  {Liebig}, {Snodgrass}, {Steele}, {Street}, \& {The RoboNet
  Collaboration}}]{Han2013}
{Han}, C., {Jung}, Y.~K., {Udalski}, A., {et~al.} 2013, \apj, 778, 38

\bibitem[{{Han} {et~al.}(2015){Han}, {Eke}, {Frenk}, {Mandelbaum}, {Norberg},
  {Schneider}, {Peacock}, {Jing}, {Baldry}, {Bland-Hawthorn}, {Brough},
  {Brown}, {Liske}, {Loveday}, \& {Robotham}}]{Han2015}
{Han}, J., {Eke}, V.~R., {Frenk}, C.~S., {et~al.} 2015, \mnras, 446, 1356

\bibitem[{{Hennawi} {et~al.}(2007){Hennawi}, {Dalal}, {Bode}, \&
  {Ostriker}}]{Hennawi2007}
{Hennawi}, J.~F., {Dalal}, N., {Bode}, P., \& {Ostriker}, J.~P. 2007, \apj,
  654, 714

\bibitem[{{Hennawi} {et~al.}(2008){Hennawi}, {Gladders}, {Oguri}, {Dalal},
  {Koester}, {Natarajan}, {Strauss}, {Inada}, {Kayo}, {Lin}, {Lampeitl},
  {Annis}, {Bahcall}, \& {Schneider}}]{Hennawi2008}
{Hennawi}, J.~F., {Gladders}, M.~D., {Oguri}, M., {et~al.} 2008, \aj, 135, 664

\bibitem[{{Hezaveh} {et~al.}(2013){Hezaveh}, {Marrone}, {Fassnacht}, {Spilker},
  {Vieira}, {Aguirre}, {Aird}, {Aravena}, {Ashby}, {Bayliss}, {Benson},
  {Bleem}, {Bothwell}, {Brodwin}, {Carlstrom}, {Chang}, {Chapman}, {Crawford},
  {Crites}, {De Breuck}, {de Haan}, {Dobbs}, {Fomalont}, {George}, {Gladders},
  {Gonzalez}, {Greve}, {Halverson}, {High}, {Holder}, {Holzapfel}, {Hoover},
  {Hrubes}, {Husband}, {Hunter}, {Keisler}, {Lee}, {Leitch}, {Lueker},
  {Luong-Van}, {Malkan}, {McIntyre}, {McMahon}, {Mehl}, {Menten}, {Meyer},
  {Mocanu}, {Murphy}, {Natoli}, {Padin}, {Plagge}, {Reichardt}, {Rest}, {Ruel},
  {Ruhl}, {Sharon}, {Schaffer}, {Shaw}, {Shirokoff}, {Stalder}, {Staniszewski},
  {Stark}, {Story}, {Vanderlinde}, {Wei{\ss}}, {Welikala}, \&
  {Williamson}}]{Hezaveh2013}
{Hezaveh}, Y.~D., {Marrone}, D.~P., {Fassnacht}, C.~D., {et~al.} 2013, \apj,
  767, 132

\bibitem[{{Hilbert} {et~al.}(2009){Hilbert}, {Hartlap}, {White}, \&
  {Schneider}}]{Hilbert2009}
{Hilbert}, S., {Hartlap}, J., {White}, S.~D.~M., \& {Schneider}, P. 2009, \aap,
  499, 31

\bibitem[{{Hilbert} {et~al.}(2007){Hilbert}, {White}, {Hartlap}, \&
  {Schneider}}]{Hilbert2007}
{Hilbert}, S., {White}, S.~D.~M., {Hartlap}, J., \& {Schneider}, P. 2007,
  \mnras, 382, 121

\bibitem[{{Hockney} \& {Eastwood}(1988)}]{Hockney1988}
{Hockney}, R.~W., \& {Eastwood}, J.~W. 1988, {Computer simulation using
  particles}

\bibitem[{{Hoekstra} \& {Jain}(2008)}]{Hoekstra2008}
{Hoekstra}, H., \& {Jain}, B. 2008, Annual Review of Nuclear and Particle
  Science, 58, 99

\bibitem[{{Illingworth} {et~al.}(2013){Illingworth}, {Magee}, {Oesch},
  {Bouwens}, {Labb{\'e}}, {Stiavelli}, {van Dokkum}, {Franx}, {Trenti},
  {Carollo}, \& {Gonzalez}}]{Illingworth2013}
{Illingworth}, G.~D., {Magee}, D., {Oesch}, P.~A., {et~al.} 2013, \apjs, 209, 6

\bibitem[{{Inoue} \& {Takahashi}(2012)}]{Inoue2012}
{Inoue}, K.~T., \& {Takahashi}, R. 2012, \mnras, 426, 2978

\bibitem[{{Jain} {et~al.}(2000){Jain}, {Seljak}, \& {White}}]{Jain2000}
{Jain}, B., {Seljak}, U., \& {White}, S. 2000, \apj, 530, 547

\bibitem[{{Jones} {et~al.}(2013){Jones}, {Ellis}, {Schenker}, \&
  {Stark}}]{Jones2013}
{Jones}, T.~A., {Ellis}, R.~S., {Schenker}, M.~A., \& {Stark}, D.~P. 2013,
  \apj, 779, 52

\bibitem[{{Jones} {et~al.}(2010){Jones}, {Swinbank}, {Ellis}, {Richard}, \&
  {Stark}}]{Jones2010}
{Jones}, T.~A., {Swinbank}, A.~M., {Ellis}, R.~S., {Richard}, J., \& {Stark},
  D.~P. 2010, \mnras, 404, 1247

\bibitem[{{Joseph} {et~al.}(2014){Joseph}, {Courbin}, {Metcalf}, {Giocoli},
  {Hartley}, {Jackson}, {Bellagamba}, {Kneib}, {Koopmans}, {Lemson},
  {Meneghetti}, {Meylan}, {Petkova}, \& {Pires}}]{Joseph2014}
{Joseph}, R., {Courbin}, F., {Metcalf}, R.~B., {et~al.} 2014, \aap, 566, A63

\bibitem[{{Keeton}(2001{\natexlab{a}})}]{Keeton2001}
{Keeton}, C.~R. 2001{\natexlab{a}}, ArXiv Astrophysics e-prints,
  astro-ph/0102341

\bibitem[{{Keeton}(2001{\natexlab{b}})}]{Keeton2001a}
---. 2001{\natexlab{b}}, \apj, 562, 160

\bibitem[{{Kelly} {et~al.}(2015){Kelly}, {Rodney}, {Treu}, {Foley}, {Brammer},
  {Schmidt}, {Zitrin}, {Sonnenfeld}, {Strolger}, {Graur}, {Filippenko}, {Jha},
  {Riess}, {Bradac}, {Weiner}, {Scolnic}, {Malkan}, {von der Linden}, {Trenti},
  {Hjorth}, {Gavazzi}, {Fontana}, {Merten}, {McCully}, {Jones}, {Postman},
  {Dressler}, {Patel}, {Cenko}, {Graham}, \& {Tucker}}]{Kelly2015}
{Kelly}, P.~L., {Rodney}, S.~A., {Treu}, T., {et~al.} 2015, Science, 347, 1123

\bibitem[{{Kilbinger} {et~al.}(2013){Kilbinger}, {Fu}, {Heymans}, {Simpson},
  {Benjamin}, {Erben}, {Harnois-D{\'e}raps}, {Hoekstra}, {Hildebrandt},
  {Kitching}, {Mellier}, {Miller}, {Van Waerbeke}, {Benabed}, {Bonnett},
  {Coupon}, {Hudson}, {Kuijken}, {Rowe}, {Schrabback}, {Semboloni}, {Vafaei},
  \& {Velander}}]{CFHT2013}
{Kilbinger}, M., {Fu}, L., {Heymans}, C., {et~al.} 2013, \mnras, 430, 2200

\bibitem[{{Killedar} {et~al.}(2012{\natexlab{a}}){Killedar}, {Borgani},
  {Meneghetti}, {Dolag}, {Fabjan}, \& {Tornatore}}]{Killedar2012b}
{Killedar}, M., {Borgani}, S., {Meneghetti}, M., {et~al.} 2012{\natexlab{a}},
  \mnras, 427, 533

\bibitem[{{Killedar} {et~al.}(2012{\natexlab{b}}){Killedar}, {Lasky}, {Lewis},
  \& {Fluke}}]{Killedar2012a}
{Killedar}, M., {Lasky}, P.~D., {Lewis}, G.~F., \& {Fluke}, C.~J.
  2012{\natexlab{b}}, \mnras, 420, 155

\bibitem[{{Kneib} \& {Natarajan}(2011)}]{Kneib2011}
{Kneib}, J.-P., \& {Natarajan}, P. 2011, \aapr, 19, 47

\bibitem[{{Komatsu} {et~al.}(2011){Komatsu}, {Smith}, {Dunkley}, {Bennett},
  {Gold}, {Hinshaw}, {Jarosik}, {Larson}, {Nolta}, {Page}, {Spergel},
  {Halpern}, {Hill}, {Kogut}, {Limon}, {Meyer}, {Odegard}, {Tucker}, {Weiland},
  {Wollack}, \& {Wright}}]{wmap7}
{Komatsu}, E., {Smith}, K.~M., {Dunkley}, J., {et~al.} 2011, \apjs, 192, 18

\bibitem[{{Kormendy} {et~al.}(2009){Kormendy}, {Fisher}, {Cornell}, \&
  {Bender}}]{Kormendy2009}
{Kormendy}, J., {Fisher}, D.~B., {Cornell}, M.~E., \& {Bender}, R. 2009, \apjs,
  182, 216

\bibitem[{{Laurikainen} {et~al.}(2010){Laurikainen}, {Salo}, {Buta}, {Knapen},
  \& {Comer{\'o}n}}]{Laurikianen2010}
{Laurikainen}, E., {Salo}, H., {Buta}, R., {Knapen}, J.~H., \& {Comer{\'o}n},
  S. 2010, \mnras, 405, 1089

\bibitem[{{Li} {et~al.}(2006){Li}, {Mao}, {Jing}, {Mo}, {Gao}, \&
  {Lin}}]{Li2006}
{Li}, G.~L., {Mao}, S., {Jing}, Y.~P., {et~al.} 2006, \mnras, 372, L73

\bibitem[{{Liao} {et~al.}(2015){Liao}, {Treu}, {Marshall}, {Fassnacht},
  {Rumbaugh}, {Dobler}, {Aghamousa}, {Bonvin}, {Courbin}, {Hojjati}, {Jackson},
  {Kashyap}, {Rathna Kumar}, {Linder}, {Mandel}, {Meng}, {Meylan}, {Moustakas},
  {Prabhu}, {Romero-Wolf}, {Shafieloo}, {Siemiginowska}, {Stalin}, {Tak},
  {Tewes}, \& {van Dyk}}]{Liao2015}
{Liao}, K., {Treu}, T., {Marshall}, P., {et~al.} 2015, \apj, 800, 11

\bibitem[{{Limousin} {et~al.}(2010){Limousin}, {Ebeling}, {Ma}, {Swinbank},
  {Smith}, {Richard}, {Edge}, {Jauzac}, {Kneib}, {Marshall}, \&
  {Schrabback}}]{Limousin2010}
{Limousin}, M., {Ebeling}, H., {Ma}, C.-J., {et~al.} 2010, \mnras, 405, 777

\bibitem[{{Limousin} {et~al.}(2012){Limousin}, {Ebeling}, {Richard},
  {Swinbank}, {Smith}, {Jauzac}, {Rodionov}, {Ma}, {Smail}, {Edge}, {Jullo}, \&
  {Kneib}}]{Limousin2012}
{Limousin}, M., {Ebeling}, H., {Richard}, J., {et~al.} 2012, \aap, 544, A71

\bibitem[{{Mahdi} {et~al.}(2014){Mahdi}, {van Beek}, {Elahi}, {Lewis}, {Power},
  \& {Killedar}}]{Mahdi2014}
{Mahdi}, H.~S., {van Beek}, M., {Elahi}, P.~J., {et~al.} 2014, \mnras, 441,
  1954

\bibitem[{{Mandelbaum} {et~al.}(2006){Mandelbaum}, {Seljak}, {Kauffmann},
  {Hirata}, \& {Brinkmann}}]{Mandelbaum2006}
{Mandelbaum}, R., {Seljak}, U., {Kauffmann}, G., {Hirata}, C.~M., \&
  {Brinkmann}, J. 2006, \mnras, 368, 715

\bibitem[{{Mandelbaum} {et~al.}(2013){Mandelbaum}, {Slosar}, {Baldauf},
  {Seljak}, {Hirata}, {Nakajima}, {Reyes}, \& {Smith}}]{Mandelbaum2013}
{Mandelbaum}, R., {Slosar}, A., {Baldauf}, T., {et~al.} 2013, \mnras, 432, 1544

\bibitem[{{Mao}(2012)}]{Mao2012}
{Mao}, S. 2012, Research in Astronomy and Astrophysics, 12, 947

\bibitem[{{Marshall} {et~al.}(2009){Marshall}, {Hogg}, {Moustakas},
  {Fassnacht}, {Brada{\v c}}, {Schrabback}, \& {Blandford}}]{Marshall2009}
{Marshall}, P.~J., {Hogg}, D.~W., {Moustakas}, L.~A., {et~al.} 2009, \apj, 694,
  924

\bibitem[{{Marshall} {et~al.}(2015){Marshall}, {Verma}, {More}, {Davis},
  {More}, {Kapadia}, {Parrish}, {Snyder}, {Wilcox}, {Baeten}, {Macmillan},
  {Cornen}, {Baumer}, {Simpson}, {Lintott}, {Miller}, {Paget}, {Simpson},
  {Smith}, {K{\"u}ng}, {Saha}, {Collett}, \& {Tecza}}]{Marshall2015}
{Marshall}, P.~J., {Verma}, A., {More}, A., {et~al.} 2015, ArXiv e-prints,
  arXiv:1504.06148

\bibitem[{{Massey} {et~al.}(2010){Massey}, {Kitching}, \&
  {Richard}}]{Massey2010}
{Massey}, R., {Kitching}, T., \& {Richard}, J. 2010, Reports on Progress in
  Physics, 73, 086901

\bibitem[{{Massey} {et~al.}(2007){Massey}, {Rhodes}, {Leauthaud}, {Capak},
  {Ellis}, {Koekemoer}, {R{\'e}fr{\'e}gier}, {Scoville}, {Taylor}, {Albert},
  {Berg{\'e}}, {Heymans}, {Johnston}, {Kneib}, {Mellier}, {Mobasher},
  {Semboloni}, {Shopbell}, {Tasca}, \& {Van Waerbeke}}]{Massey2007}
{Massey}, R., {Rhodes}, J., {Leauthaud}, A., {et~al.} 2007, \apjs, 172, 239

\bibitem[{{McCully} {et~al.}(2014){McCully}, {Keeton}, {Wong}, \&
  {Zabludoff}}]{McCully2014}
{McCully}, C., {Keeton}, C.~R., {Wong}, K.~C., \& {Zabludoff}, A.~I. 2014,
  \mnras, 443, 3631

\bibitem[{{McLeod} {et~al.}(2015){McLeod}, {Geary}, {Conroy}, {Fabricant},
  {Ordway}, {Szentgyorgyi}, {Amato}, {Ashby}, {Caldwell}, {Curley}, {Gauron},
  {Holman}, {Norton}, {Pieri}, {Roll}, {Weaver}, {Zajac}, {Palunas}, \&
  {Osip}}]{mcleod06}
{McLeod}, B., {Geary}, J., {Conroy}, M., {et~al.} 2015, \pasp, 127, 366

\bibitem[{{Meneghetti} {et~al.}(2007){Meneghetti}, {Argazzi}, {Pace},
  {Moscardini}, {Dolag}, {Bartelmann}, {Li}, \& {Oguri}}]{Meneghetti2007}
{Meneghetti}, M., {Argazzi}, R., {Pace}, F., {et~al.} 2007, \aap, 461, 25

\bibitem[{{Meneghetti} {et~al.}(2013){Meneghetti}, {Bartelmann}, {Dahle}, \&
  {Limousin}}]{Meneghetti2013}
{Meneghetti}, M., {Bartelmann}, M., {Dahle}, H., \& {Limousin}, M. 2013, \ssr,
  177, 31

\bibitem[{{Meneghetti} {et~al.}(2003{\natexlab{a}}){Meneghetti}, {Bartelmann},
  \& {Moscardini}}]{Meneghetti2003a}
{Meneghetti}, M., {Bartelmann}, M., \& {Moscardini}, L. 2003{\natexlab{a}},
  \mnras, 346, 67

\bibitem[{{Meneghetti} {et~al.}(2003{\natexlab{b}}){Meneghetti}, {Bartelmann},
  \& {Moscardini}}]{Meneghetti2003b}
---. 2003{\natexlab{b}}, \mnras, 340, 105

\bibitem[{{Meneghetti} {et~al.}(2010){Meneghetti}, {Rasia}, {Merten},
  {Bellagamba}, {Ettori}, {Mazzotta}, {Dolag}, \& {Marri}}]{Meneghetti2010}
{Meneghetti}, M., {Rasia}, E., {Merten}, J., {et~al.} 2010, \aap, 514, A93

\bibitem[{{Meneghetti} {et~al.}(2008){Meneghetti}, {Melchior}, {Grazian}, {De
  Lucia}, {Dolag}, {Bartelmann}, {Heymans}, {Moscardini}, \&
  {Radovich}}]{Meneghetti2008}
{Meneghetti}, M., {Melchior}, P., {Grazian}, A., {et~al.} 2008, \aap, 482, 403

\bibitem[{{Meneghetti} {et~al.}(2014){Meneghetti}, {Rasia}, {Vega}, {Merten},
  {Postman}, {Yepes}, {Sembolini}, {Donahue}, {Ettori}, {Umetsu}, {Balestra},
  {Bartelmann}, {Ben{\'{\i}}tez}, {Biviano}, {Bouwens}, {Bradley},
  {Broadhurst}, {Coe}, {Czakon}, {De Petris}, {Ford}, {Giocoli},
  {Gottl{\"o}ber}, {Grillo}, {Infante}, {Jouvel}, {Kelson}, {Koekemoer},
  {Lahav}, {Lemze}, {Medezinski}, {Melchior}, {Mercurio}, {Molino},
  {Moscardini}, {Monna}, {Moustakas}, {Moustakas}, {Nonino}, {Rhodes},
  {Rosati}, {Sayers}, {Seitz}, {Zheng}, \& {Zitrin}}]{Meneghetti2014}
{Meneghetti}, M., {Rasia}, E., {Vega}, J., {et~al.} 2014, \apj, 797, 34

\bibitem[{{Metcalf} \& {Petkova}(2014)}]{Metcalf2014}
{Metcalf}, R.~B., \& {Petkova}, M. 2014, \mnras, 445, 1942

\bibitem[{{Moffat}(1969)}]{Moffat1969}
{Moffat}, A.~F.~J. 1969, \aap, 3, 455

\bibitem[{{More} {et~al.}(2012){More}, {Cabanac}, {More}, {Alard}, {Limousin},
  {Kneib}, {Gavazzi}, \& {Motta}}]{More2012}
{More}, A., {Cabanac}, R., {More}, S., {et~al.} 2012, \apj, 749, 38

\bibitem[{{More} {et~al.}(2016){More}, {Verma}, {Marshall}, {More}, {Baeten},
  {Wilcox}, {Macmillan}, {Cornen}, {Kapadia}, {Parrish}, {Snyder}, {Davis},
  {Gavazzi}, {Lintott}, {Simpson}, {Miller}, {Smith}, {Paget}, {Saha},
  {K{\"u}ng}, \& {Collett}}]{More2016}
{More}, A., {Verma}, A., {Marshall}, P.~J., {et~al.} 2016, \mnras, 455, 1191

\bibitem[{{Muraki} {et~al.}(2011){Muraki}, {Han}, {Bennett}, {Suzuki},
  {Monard}, {Street}, {Jorgensen}, {Kundurthy}, {Skowron}, {Becker}, {Albrow},
  {Fouqu{\'e}}, {Heyrovsk{\'y}}, {Barry}, {Beaulieu}, {Wellnitz}, {Bond},
  {Sumi}, {Dong}, {Gaudi}, {Bramich}, {Dominik}, {Abe}, {Botzler}, {Freeman},
  {Fukui}, {Furusawa}, {Hayashi}, {Hearnshaw}, {Hosaka}, {Itow}, {Kamiya},
  {Korpela}, {Kilmartin}, {Lin}, {Ling}, {Makita}, {Masuda}, {Matsubara},
  {Miyake}, {Nishimoto}, {Ohnishi}, {Perrott}, {Rattenbury}, {Saito},
  {Skuljan}, {Sullivan}, {Sweatman}, {Tristram}, {Wada}, {Yock}, {MOA
  Collaboration}, {Christie}, {DePoy}, {Gorbikov}, {Gould}, {Kaspi}, {Lee},
  {Mallia}, {Maoz}, {McCormick}, {Moorhouse}, {Natusch}, {Park}, {Pogge},
  {Polishook}, {Shporer}, {Thornley}, {Yee}, {{$\mu$}FUN Collaboration},
  {Allan}, {Browne}, {Horne}, {Kains}, {Snodgrass}, {Steele}, {Tsapras},
  {RoboNet Collaboration}, {Batista}, {Bennett}, {Brillant}, {Caldwell},
  {Cassan}, {Cole}, {Corrales}, {Coutures}, {Dieters}, {Dominis Prester},
  {Donatowicz}, {Greenhill}, {Kubas}, {Marquette}, {Martin}, {Menzies}, {Sahu},
  {Waldman}, {Williams}, {Zub}, {PLANET Collaboration}, {Bourhrous},
  {Matsuoka}, {Nagayama}, {Oi}, {Randriamanakoto}, {IRSF Observers}, {Bozza},
  {Burgdorf}, {Calchi Novati}, {Dreizler}, {Finet}, {Glitrup}, {Harps{\o}e},
  {Hinse}, {Hundertmark}, {Liebig}, {Maier}, {Mancini}, {Mathiasen}, {Rahvar},
  {Ricci}, {Scarpetta}, {Skottfelt}, {Surdej}, {Southworth}, {Wambsganss},
  {Zimmer}, {MiNDSTEp Consortium}, {Udalski}, {Poleski}, {Wyrzykowski},
  {Ulaczyk}, {Szyma{\'n}ski}, {Kubiak}, {Pietrzy{\'n}ski}, {Soszy{\'n}ski}, \&
  {OGLE Collaboration}}]{Muraki2011}
{Muraki}, Y., {Han}, C., {Bennett}, D.~P., {et~al.} 2011, \apj, 741, 22

\bibitem[{{Murphy} {et~al.}(2011){Murphy}, {Gebhardt}, \& {Adams}}]{Murphy2011}
{Murphy}, J.~D., {Gebhardt}, K., \& {Adams}, J.~J. 2011, \apj, 729, 129

\bibitem[{{Nagai} \& {Kravtsov}(2005)}]{nagai2005}
{Nagai}, D., \& {Kravtsov}, A.~V. 2005, \apj, 618, 557

\bibitem[{{Narayan} \& {Bartelmann}(1996)}]{Narayan1996}
{Narayan}, R., \& {Bartelmann}, M. 1996, ArXiv Astrophysics e-prints,
  astro-ph/9606001

\bibitem[{{Navarro} {et~al.}(1996){Navarro}, {Frenk}, \&
  {White}}]{1996ApJ...462..563N}
{Navarro}, J.~F., {Frenk}, C.~S., \& {White}, S.~D.~M. 1996, \apj, 462, 563

\bibitem[{{Newman} {et~al.}(2011){Newman}, {Treu}, {Ellis}, \&
  {Sand}}]{Newman2011}
{Newman}, A.~B., {Treu}, T., {Ellis}, R.~S., \& {Sand}, D.~J. 2011, \apjl, 728,
  L39

\bibitem[{{Newman} {et~al.}(2013){Newman}, {Treu}, {Ellis}, \&
  {Sand}}]{Newman2013}
---. 2013, \apj, 765, 25

\bibitem[{{Newman} {et~al.}(2009){Newman}, {Treu}, {Ellis}, {Sand}, {Richard},
  {Marshall}, {Capak}, \& {Miyazaki}}]{Newman2009}
{Newman}, A.~B., {Treu}, T., {Ellis}, R.~S., {et~al.} 2009, \apj, 706, 1078

\bibitem[{{Oguri} {et~al.}(2012){Oguri}, {Bayliss}, {Dahle}, {Sharon},
  {Gladders}, {Natarajan}, {Hennawi}, \& {Koester}}]{Oguri2012}
{Oguri}, M., {Bayliss}, M.~B., {Dahle}, H., {et~al.} 2012, \mnras, 420, 3213

\bibitem[{{Oguri} \& {Blandford}(2009)}]{Oguri2009}
{Oguri}, M., \& {Blandford}, R.~D. 2009, \mnras, 392, 930

\bibitem[{{Okabe} {et~al.}(2010){Okabe}, {Takada}, {Umetsu}, {Futamase}, \&
  {Smith}}]{Okabe2010}
{Okabe}, N., {Takada}, M., {Umetsu}, K., {Futamase}, T., \& {Smith}, G.~P.
  2010, \pasj, 62, 811

\bibitem[{{Paraficz} \& {Hjorth}(2010)}]{Paraficz2010}
{Paraficz}, D., \& {Hjorth}, J. 2010, \apj, 712, 1378

\bibitem[{{Peter} {et~al.}(2013){Peter}, {Rocha}, {Bullock}, \&
  {Kaplinghat}}]{Peter2013}
{Peter}, A.~H.~G., {Rocha}, M., {Bullock}, J.~S., \& {Kaplinghat}, M. 2013,
  \mnras, 430, 105

\bibitem[{{Petkova} {et~al.}(2014){Petkova}, {Metcalf}, \&
  {Giocoli}}]{Petkova2014}
{Petkova}, M., {Metcalf}, R.~B., \& {Giocoli}, C. 2014, \mnras, 445, 1954

\bibitem[{{Puchwein} \& {Hilbert}(2009)}]{Puchwein2009}
{Puchwein}, E., \& {Hilbert}, S. 2009, \mnras, 398, 1298

\bibitem[{{Rasia} {et~al.}(2012){Rasia}, {Meneghetti}, {Martino}, {Borgani},
  {Bonafede}, {Dolag}, {Ettori}, {Fabjan}, {Giocoli}, {Mazzotta}, {Merten},
  {Radovich}, \& {Tornatore}}]{Rasia2012}
{Rasia}, E., {Meneghetti}, M., {Martino}, R., {et~al.} 2012, New Journal of
  Physics, 14, 055018

\bibitem[{{Richard} {et~al.}(2008){Richard}, {Stark}, {Ellis}, {George},
  {Egami}, {Kneib}, \& {Smith}}]{Richard2008}
{Richard}, J., {Stark}, D.~P., {Ellis}, R.~S., {et~al.} 2008, \apj, 685, 705

\bibitem[{{Rowe} {et~al.}(2014){Rowe}, {Jarvis}, {Mandelbaum}, {Bernstein},
  {Bosch}, {Simet}, {Meyers}, {Kacprzak}, {Nakajima}, {Zuntz}, {Miyatake},
  {Dietrich}, {Armstrong}, {Melchior}, \& {Gill}}]{Rowe2014}
{Rowe}, B., {Jarvis}, M., {Mandelbaum}, R., {et~al.} 2014, ArXiv e-prints,
  arXiv:1407.7676

\bibitem[{{Rozo} {et~al.}(2008){Rozo}, {Nagai}, {Keeton}, \&
  {Kravtsov}}]{Rozo2008}
{Rozo}, E., {Nagai}, D., {Keeton}, C., \& {Kravtsov}, A. 2008, \apj, 687, 22

\bibitem[{{Ryden}(1992)}]{Ryden1992}
{Ryden}, B. 1992, \apj, 396, 445

\bibitem[{{Saha}(2000)}]{Saha2000}
{Saha}, P. 2000, \aj, 120, 1654

\bibitem[{{Sand} {et~al.}(2008){Sand}, {Treu}, {Ellis}, {Smith}, \&
  {Kneib}}]{Sand2008}
{Sand}, D.~J., {Treu}, T., {Ellis}, R.~S., {Smith}, G.~P., \& {Kneib}, J.-P.
  2008, \apj, 674, 711

\bibitem[{{Schaap} \& {van de Weygaert}(2000)}]{Schaap2000}
{Schaap}, W.~E., \& {van de Weygaert}, R. 2000, \aap, 363, L29

\bibitem[{{Schneider}(2014)}]{Scheinder2014}
{Schneider}, P. 2014, \aap, 568, L2

\bibitem[{{Schneider} {et~al.}(1992){Schneider}, {Ehlers}, \&
  {Falco}}]{Schneider1992}
{Schneider}, P., {Ehlers}, J., \& {Falco}, E.~E. 1992, {Gravitational Lenses},
  ed. {Schneider, P., Ehlers, J., \& Falco, E.~E.}

\bibitem[{{Seidel} \& {Bartelmann}(2007)}]{Seidel2007}
{Seidel}, G., \& {Bartelmann}, M. 2007, \aap, 472, 341

\bibitem[{{Sereno} \& {Covone}(2013)}]{Sereno2013}
{Sereno}, M., \& {Covone}, G. 2013, \mnras, 434, 878

\bibitem[{{Springel} {et~al.}(2001){Springel}, {White}, {Tormen}, \&
  {Kauffmann}}]{subfind}
{Springel}, V., {White}, S.~D.~M., {Tormen}, G., \& {Kauffmann}, G. 2001,
  \mnras, 328, 726

\bibitem[{{Stark} {et~al.}(2013){Stark}, {Auger}, {Belokurov}, {Jones},
  {Robertson}, {Ellis}, {Sand}, {Moiseev}, {Eagle}, \& {Myers}}]{Stark2013}
{Stark}, D.~P., {Auger}, M., {Belokurov}, V., {et~al.} 2013, \mnras, 436, 1040

\bibitem[{{Stark} {et~al.}(2014){Stark}, {Richard}, {Charlot}, {Clement},
  {Ellis}, {Siana}, {Robertson}, {Schenker}, {Gutkin}, \&
  {Wofford}}]{Stark2014}
{Stark}, D.~P., {Richard}, J., {Charlot}, S., {et~al.} 2014, ArXiv e-prints,
  arXiv:1408.3649

\bibitem[{{Suyu} {et~al.}(2010){Suyu}, {Marshall}, {Auger}, {Hilbert},
  {Blandford}, {Koopmans}, {Fassnacht}, \& {Treu}}]{Suyu2010}
{Suyu}, S.~H., {Marshall}, P.~J., {Auger}, M.~W., {et~al.} 2010, \apj, 711, 201

\bibitem[{{Suyu} {et~al.}(2013){Suyu}, {Auger}, {Hilbert}, {Marshall}, {Tewes},
  {Treu}, {Fassnacht}, {Koopmans}, {Sluse}, {Blandford}, {Courbin}, \&
  {Meylan}}]{Suyu2013}
{Suyu}, S.~H., {Auger}, M.~W., {Hilbert}, S., {et~al.} 2013, \apj, 766, 70

\bibitem[{{Suyu} {et~al.}(2014){Suyu}, {Treu}, {Hilbert}, {Sonnenfeld},
  {Auger}, {Blandford}, {Collett}, {Courbin}, {Fassnacht}, {Koopmans},
  {Marshall}, {Meylan}, {Spiniello}, \& {Tewes}}]{Suyu2014}
{Suyu}, S.~H., {Treu}, T., {Hilbert}, S., {et~al.} 2014, \apjl, 788, L35

\bibitem[{{Takahashi} {et~al.}(2011){Takahashi}, {Oguri}, {Sato}, \&
  {Hamana}}]{Takahashi2011}
{Takahashi}, R., {Oguri}, M., {Sato}, M., \& {Hamana}, T. 2011, \apj, 742, 15

\bibitem[{{Treu}(2010)}]{Treu2010}
{Treu}, T. 2010, \araa, 48, 87

\bibitem[{{Treu} {et~al.}(2013){Treu}, {Marshall}, {Cyr-Racine}, {Fassnacht},
  {Keeton}, {Linder}, {Moustakas}, {Bradac}, {Buckley-Geer}, {Collett},
  {Courbin}, {Dobler}, {Finley}, {Hjorth}, {Kochanek}, {Komatsu}, {Koopmans},
  {Meylan}, {Natarajan}, {Oguri}, {Suyu}, {Tewes}, {Wong}, {Zabludoff},
  {Zaritsky}, {Anguita}, {Brunner}, {Cabanac}, {Falco}, {Fritz}, {Seidel},
  {Howell}, {Giocoli}, {Jackson}, {Lopez}, {Metcalf}, {Motta}, \&
  {Verdugo}}]{Treu2013}
{Treu}, T., {Marshall}, P.~J., {Cyr-Racine}, F.-Y., {et~al.} 2013, ArXiv
  e-prints, arXiv:1306.1272

\bibitem[{{Umetsu} \& {Broadhurst}(2008)}]{Umetsu2008}
{Umetsu}, K., \& {Broadhurst}, T. 2008, \apj, 684, 177

\bibitem[{{Vale} \& {White}(2003)}]{Vale2003}
{Vale}, C., \& {White}, M. 2003, \apj, 592, 699

\bibitem[{{van Engelen} {et~al.}(2012){van Engelen}, {Keisler}, {Zahn}, {Aird},
  {Benson}, {Bleem}, {Carlstrom}, {Chang}, {Cho}, {Crawford}, {Crites}, {de
  Haan}, {Dobbs}, {Dudley}, {George}, {Halverson}, {Holder}, {Holzapfel},
  {Hoover}, {Hou}, {Hrubes}, {Joy}, {Knox}, {Lee}, {Leitch}, {Lueker},
  {Luong-Van}, {McMahon}, {Mehl}, {Meyer}, {Millea}, {Mohr}, {Montroy},
  {Natoli}, {Padin}, {Plagge}, {Pryke}, {Reichardt}, {Ruhl}, {Sayre},
  {Schaffer}, {Shaw}, {Shirokoff}, {Spieler}, {Staniszewski}, {Stark}, {Story},
  {Vanderlinde}, {Vieira}, \& {Williamson}}]{Engelen2012}
{van Engelen}, A., {Keisler}, R., {Zahn}, O., {et~al.} 2012, \apj, 756, 142

\bibitem[{{Wambsganss} {et~al.}(2004){Wambsganss}, {Bode}, \&
  {Ostriker}}]{Wambsganss2004}
{Wambsganss}, J., {Bode}, P., \& {Ostriker}, J.~P. 2004, \apjl, 606, L93

\bibitem[{{Wambsganss} {et~al.}(2005){Wambsganss}, {Bode}, \&
  {Ostriker}}]{Wambsganss2005}
---. 2005, \apjl, 635, L1

\bibitem[{{Wen} {et~al.}(2011){Wen}, {Han}, \& {Jiang}}]{Wen2011}
{Wen}, Z.-L., {Han}, J.-L., \& {Jiang}, Y.-Y. 2011, Research in Astronomy and
  Astrophysics, 11, 1185

\bibitem[{{Wuyts} {et~al.}(2012){Wuyts}, {Rigby}, {Gladders}, {Gilbank},
  {Sharon}, {Gralla}, \& {Bayliss}}]{Wuyts2012}
{Wuyts}, E., {Rigby}, J.~R., {Gladders}, M.~D., {et~al.} 2012, \apj, 745, 86

\bibitem[{{Xu} {et~al.}(2016){Xu}, {Postman}, {Meneghetti}, {Seitz}, {Zitrin},
  {Merten}, {Maoz}, {Frye}, {Umetsu}, {Zheng}, {Bradley}, {Vega}, \&
  {Koekemoer}}]{Xu2016}
{Xu}, B., {Postman}, M., {Meneghetti}, M., {et~al.} 2016, \apj, 817, 85

\bibitem[{{Xu} {et~al.}(2009){Xu}, {Mao}, {Wang}, {Springel}, {Gao}, {White},
  {Frenk}, {Jenkins}, {Li}, \& {Navarro}}]{Xu2009}
{Xu}, D.~D., {Mao}, S., {Wang}, J., {et~al.} 2009, \mnras, 398, 1235

\end{thebibliography}

\clearpage

\end{document}